\documentclass[12pt]{iopart}
\pdfoutput=1

\usepackage{iopams,hyperref}
\usepackage{graphicx, xcolor}  

\usepackage{braket}
\usepackage{verbatim}
\usepackage{cite}

\begin{document}
\title[A finite temperature study of ideal quantum gases in 1D quasi-periodic potential]{A finite temperature study of ideal quantum gases in the presence of one dimensional quasi-periodic potential}

\author{Nilanjan Roy$^1$ and Subhasis Sinha$^2$}

\address{$^1$ \it Department of Physics, Indian Institute of Science Education and Research, Bhopal, Madhya Pradesh 462066, India}
\address{$^2$ \it Indian Institute of Science Education and Research-Kolkata, Mohanpur, Nadia 741246, India}



\begin{abstract}
We study the thermodynamics of ideal Bose gas as well as the transport properties of non interacting bosons and fermions in a one dimensional quasi-periodic potential, namely Aubry-Andr\'e (AA) model at finite temperature. 
For bosons in finite size systems, the effect of quasi-periodic potential on the crossover phenomena corresponding to Bose-Einstein condensation (BEC), superfluidity and localization phenomena at finite temperatures are investigated. 
From the ground state number fluctuation we calculate the crossover temperature of BEC which exhibits a non monotonic behavior with the strength of AA potential and vanishes at the self-dual critical point following power law. Appropriate rescaling of the crossover temperatures reveals universal behavior which is studied for different quasi-periodicity of the AA model.
Finally, we study the temperature and flux dependence of the persistent current of fermions in presence of a quasi-periodic potential to identify the localization at the Fermi energy from the decay of the current. 
\end{abstract}
\noindent{\it Keywords\/}:: Quantum gases, Quantum Disordered Systems, Bose Einstein condensation, Quantum transport in one-dimension

\submitto{\JSTAT} 
\maketitle
\section{Introduction}
Over the past few decades much effort has been devoted to understand the properties of quantum system in the presence of a quasi-periodic potential
\cite{Goldman,kohomoto1,kohomoto2,physrep,repprogphys}.
In recent years the study of quasi-periodic system has regained interest in the context of localization phenomena. Unlike Anderson model \cite{Anderson}, an incommensurate lattice potential which is known as Aubry-Andr\'e (AA) model can exhibit a localization transition in one dimension\cite{AA}. 
In recent seminal experiments, it has been demonstrated that such incommensurate quasi-periodic potential can indeed localize light\cite{light} and ultracold matter waves\cite{inguscio0,inguscio1,inguscio2}.
This experimental observation has generated an impetus to study the effect of interaction on localization of dilute condensate\cite{inguscio3,inguscio4,palencia,sinha} as well as `many body localization' phenomena\cite{rev1,huse} and correlated glassy phases\cite{glass1,glass2,glass3,glass4,minguzzi1,minguzzi2,krishnendu} in the strongly interacting many particle systems. Also the experimental realization of AA-model has provided test-bed to study many body localization-delocalization transition\cite{mbl_bloch} and dynamical delocalization phenomena induced by periodic drive\cite{drive,sayak}.

Fate of the many body localized phases under thermal fluctuation and localization transition at finite temperatures are the emergent issues which require further investigations. In presence of random impurities the metal-insulator transition(MIT) at finite temperatures has been studied for electronic systems\cite{Imry}. Recent theoretical works predict a non-conventional finite temperature transition from localized phase to a quantum fluid phase in one dimension\cite{slyap,michal}. The phase boundary of such insulator-fluid transition is predicted for weakly interacting Bose gas in quasi-periodic potential\cite{michal}. A signature of such finite temperature effect has also been found in localized phase of hardcore bosons\cite{nessi}. The finite temperature phases and localization transition of strongly as well as weakly interacting bosons and fermions in the presence of disorder deserves further study.

To gain a better insight of localization transition at finite temperature, as a first step it is important to study the thermodynamic and transport properties of ideal quantum gases in the presence of a 1D quasi-periodic potential which can also guide us to identify possible new phases and effects arising from interactions. 
Even in absence of interaction the AA model has relevance because of its properties like `duality' and fractal behavior of energy spectrum.  
Recent studies revealed non-trivial behavior of certain thermodynamic properties of quasi-periodic systems because of multi-scale fractal structure of energy spectrum\cite{Tsalis}.  
Moreover, the duality of the AA model is an important aspect due to which the localization transition occurs at the self-dual critical point in absence of mobility edge\cite{AA}. 
Similar to quantum phase transition, manifestation of the self-dual critical point at finite temperature is a relevant issue related to the thermodynamic behavior of non-interacting AA model. 
After the achievement of Bose-Einstein condensation in trapped dilute gases, the study of thermodynamic properties of finite systems revealed several interesting aspects even in absence of interaction, related to the finite size effects, fluctuation phenomena and choice of appropriate ensemble for such isolated finite quantum systems\cite{ketterle,weiss,bec_fluc,tran}. Investigation of these aspects of AA model at finite temperature is also important for ongoing experiments on ultracold quantum gases in bichromatic optical lattice where tunability of the interaction strength enables to study ideal quantum gases in quasi-periodic lattice\cite{inguscio1}.
 
In this work, we investigate the finite temperature thermodynamics of non-interacting bosons and fermions in the AA-potential focusing on the localization transition.
In the first part, the influence of quasi-periodic disorder on the formation of `quasi-condensate' and superfluidity of 1D Bose gas at finite temperature are studied. Although in one dimension a true transition at finite temperature is absent in the thermodynamic limit\cite{pathria}, a crossover phenomena to the Bose-Einstein condensate (BEC) phase can occur in finite size system\cite{ketterle,shlyapnikov,shlyap_rev}.
From the ground state number fluctuation we calculate the crossover temperature corresponding to BEC in both canonical and grand canonical ensemble to study its dependence on both strength and quasi-periodicity of the AA potential, effect of self-dual critical point and ensemble equivalence. The transport and localization phenomena are investigated from finite temperature superfluid fraction and temperature dependent participation ratio. 
%

Similar to the superfluid fraction the persistent current is another physical quantity which measures the transport properties of fermionic system. The persistent current in mesoscopic quantum ring in the presence of a magnetic flux and its variation with the physical parameters like magnetic flux and temperature are well studied subjects\cite{Imry}. The decay of the persistent current in presence of disorder\cite{cheung1,bouchiat} and aperiodic potential\cite{thumorse,fibonacci} is also studied. To investigate the localization of fermionic system in AA potential we study the persistent current at finite temperatures. 

The paper is organized as follows: in section \ref{sec2} we review the Aubry-Andr\'e model and single particle localization transition due to the self duality of the model. Thermodynamics of non-interacting bosons in the AA potential is presented in section \ref{sec3}. In subsection \ref{sec3.1} we investigate the crossover to condensate phase and from the ground state number fluctuations calculate the crossover temperature which reveals the signature of self dual critical point at finite temperature.
The superfluid fraction (SFF) and inverse participation ratio (IPR) are presented in subsection \ref{sec3.2} to study the localization phenomena at finite temperature.
In subsection \ref{sec3.3}, we calculate the single particle entanglement entropy at finite temperatures. Apart from localization transition, we identify different phases of the Bose gas at finite temperature from these thermodynamic quantities.
In section \ref{sec4}, we study the persistent current of non-interacting fermions at finite temperature to capture the localization transition near the Fermi energy. 
The persistent current at zero temperature and at finite temperature are discussed separately in subsection \ref{sec4.1} and subsection \ref{sec4.2}.
Finally, we summarize our results and conclude in section \ref{sec5}.

\section{ Localization transition in Aubry-Andr\'e model}\label{sec2}
In this section we consider an incommensurate lattice potential in one dimension which is known as Aubry-Andr\'e model\cite{AA}. This particular model has drawn much interest since it exhibits the localization transition in one dimension due to its self-dual property\cite{AA,ingold,larcher1,larcher2}. In the recent experiments with ultracold atoms, the AA model has been engineered by using bichromatic optical lattice \cite{inguscio0,modugno}. The Hamiltonian of the AA model is given by,
\begin{eqnarray}
H = -J \sum\limits_{l=1}^{N_s}(a_{l+1}^{\dagger}a_{l} + h.c.) +  \lambda \sum\limits_l \cos(2\pi\alpha l) n_l,
\label{eq1} 
\end{eqnarray}
where $a_{l}$($a_{l}^{\dagger}$) is single-particle annihilation(creation) operator at site $l$, $n_l=a_{l}^{\dagger}a_{l}$ is the number operator, $J$ is the nearest-neighbor hopping strength, $\lambda$ is the strength of the potential. Here we consider a lattice of length $L=N_s \bar{a}$ with $N_s$ number of sites and lattice spacing $\bar{a}$. The quasi-periodic nature of the potential is generated by choosing the parameter $\alpha$ to be an irrational number.
Localization transition in AA model is essentially a single particle phenomena determined from the eigenfunctions of the above Hamiltonian.
It can be shown that when $\alpha$ is taken to be a Diophantine number, all the eigenstates become localized above a critical strength $\lambda=2J$ \cite{mathieu}. Any irrational number can be written as continued faction\cite{continuedfraction} which allows successive rational approximation of it in the form of $p/q$ where $p$, $q$ are coprime numbers. For Diophantine numbers the accuracy of the rational approximation has a lower bound, such that $|\alpha-\frac{p}{q}|> c/q^{2+\zeta}$ with $c>0$ and $\zeta\geq 0$ \cite{diophantine}.
A sequence of Diophantine numbers known as `metallic means' can be obtained from the generalized Fibonacci series, 
\begin{equation}
F_{p}=s_1 F_{p-1} + F_{p-2},
\label{diophantine}
\end{equation}
with $F_0=0, F_1=1$, also known as the `Metallic Mean Family'\cite{mmf}. From this series an approximation of the Diophantine number is given by $\alpha=F_{p-1}/F_p$ for successive value of the integer $p$ and converges to the value of `metallic mean' in the limit of $p \rightarrow \infty$. 
For successive integer values of $s_1 =1,2,3,4.....$ the sequence of Diophantine numbers generated from the above series are generally known as `golden', `silver',`bronze' and `copper' means which are given by $\alpha_g= (\sqrt{5}-1)/2$, $\alpha_s= \sqrt{2}-1$, $\alpha_b=  (\sqrt{13}-3)/2 $ and $\alpha_c= \sqrt{5} - 2$ respectively.
  
In the localization transition of the AA model `duality' plays a crucial role, which can be shown from the transformations in the `momentum' space,
$a_l=\frac{1}{\sqrt{N_s}} \sum_{k} a_k e^{i(2\pi\alpha k + \pi)l}$. 
The transformed Hamiltonian in the momentum space is written as,   
\begin{eqnarray}
H(k)=-\frac{\lambda}{2} \sum\limits_k (a_{k+1}^{\dagger}a_k+h.c.)+2J \sum\limits_k \cos(2\pi\alpha k)n_k.
\end{eqnarray} 
It is evident that the transformed Hamiltonian is exactly same as that of the original AA model with modified coupling constants. For $\lambda=2J$, both the Hamiltonians in momentum space and real space become identical to each other exhibiting `self-duality' of the AA model at this critical coupling strength. As a consequence of the self-duality, all eigenstates become localized at the critical coupling $\lambda_c =2J$ with energy independent localization length \cite{thoules}. The localization transition without mobility edge at the critical coupling strength is an important property which makes the AA model different from the Anderson impurity model \cite{Anderson}. In this work we investigate the influence of the `self-dual' critical point in the finite temperature thermodynamic quantities of non-interacting bosons and fermions.

In the numerical calculations, we impose the periodic boundary condition by making a rational approximation of the parameter $\alpha=\frac{F_{p-1}}{F_p}$ and by choosing the lattice site $N_s=F_p$ \cite{ingold}. 
Here $F_p$ is the $p$th term in the Fibonacci series for sufficiently large $p$ so that the rational approximation of the `metallic mean' becomes valid in the thermodynamic limit. In the rest of the paper, we denote all energy scales and temperatures $kT$ in the unit $J$ and use $J=1$. For $N$ particle in a lattice of total $N_s$ sites the filling factor is given by $\nu = N/N_s$. We consider a lattice of total sites $N_s=360$ and $1292$ for quasi-periodicity $\alpha=\alpha_b$ and $\alpha_c$ respectively, unless specifically mentioned for all the figures.  
 
To quantify the degree of localization of single particle wavefunctions, we calculate the inverse participation ratio (IPR) which is defined as,
\begin{equation}
I = \sum_{l} {|\phi_l|}^4
\label{ipr},
\end{equation}
where $\phi_l$ is the amplitude of the normalized wavefunction at site $l$. For extended states the IPR vanishes in the thermodynamic limit, whereas it saturates to unity for wavefunction localized at a single site.
\begin{figure}
\centering
\includegraphics[width=5.4cm,height=4.7cm]{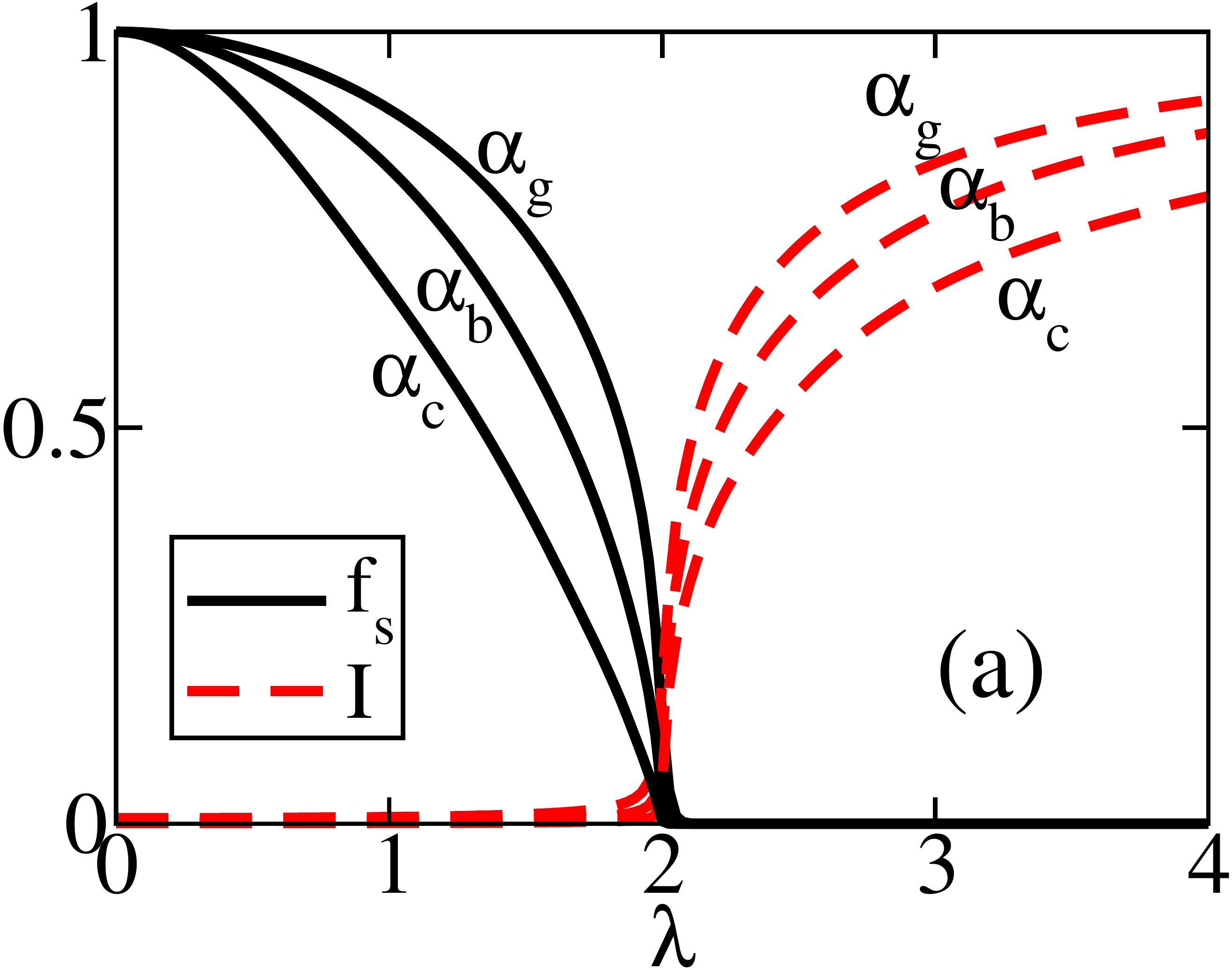}~
 \includegraphics[width=6.0cm,height=5.4cm]{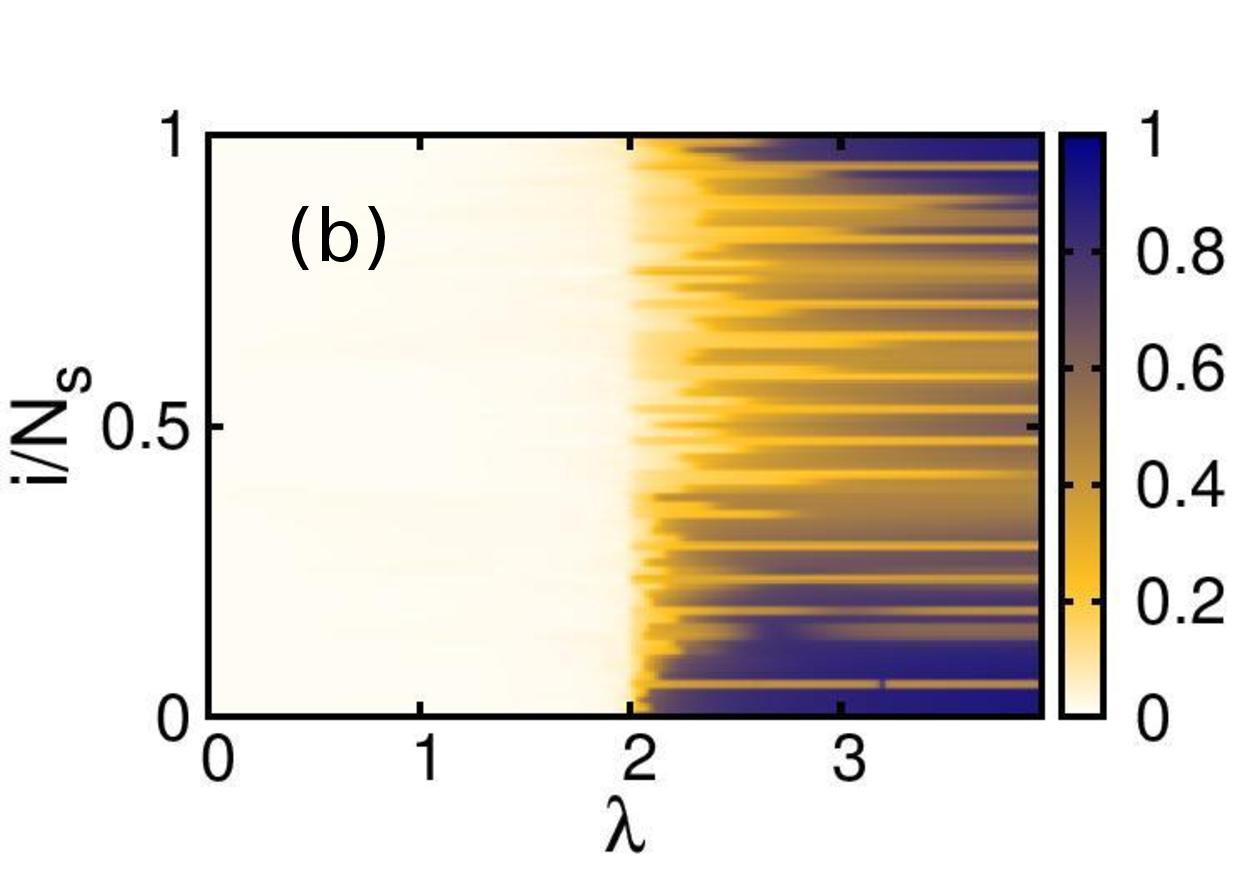}
\caption{(a) Variation of the SFF $f_s$ at zero temperature and IPR of the ground state $I$ with the strength of the quasi-periodic potential $\lambda$ for increasing values of $\alpha$. (b) Inverse participation ratio (IPR) of all the single particle eigenstates as a function of $\lambda$ for $\alpha=\alpha_g$. Here eigenstates are denoted by the index $i$ in the ascending order of energies. For both the plots $N_s=144$ for $\alpha=\alpha_g$.}
\label{zeroT}
\end{figure}

More robust signature of localization in a physical system can be observed from the change in the transport properties, which vanishes in the localized regime. 
The `superfluid fraction' (SFF) is a natural quantity to study superfluid transport in a neutral bosonic system, similar to the conductivity in electronic systems. A direct detection of localization transition of the single particle ground state of AA Hamiltonian can be done from SFF since it is occupied by all non interacting bosons at zero temperature.
The SFF is measured from a superflow of bosons which is generated by applying a phase twist $\theta$ in the boundary \cite{fisher}. The phase-twisted Hamiltonian in one dimension is identical to the original Hamiltonian with modified hopping strength $J_{l,l^\prime} =Je^{i\theta(l^\prime-l)/N_s}$ between site ${l^\prime}$ to site $l$. In presence of a total twist $\theta$ at the boundary, the Hamiltonian is given by,
\begin{eqnarray}
H (\theta) = \sum_{\langle l,l^\prime \rangle}( -Je^{-i\theta/N_s} a_{l}^{\dagger}a_{l^\prime} + h.c.) +  \lambda\sum\limits_l \cos(2\pi\alpha l) n_l,
\label{ham_twist}
\end{eqnarray}
where $\langle l,l^\prime \rangle$ denotes the nearest neighbor lattice sites. At zero temperature and for small phase twist $\theta \ll \pi$, the SFF is defined as \cite{fisher,roth}, 
\begin{eqnarray}
&& f_s=\frac{{N_s}^2}{N} \frac{E(\theta) - E(0)}{{\theta}^2}, 
\end{eqnarray}
where $E(\theta)$ is the ground state energy of the Hamiltonian with twist $\theta$. 
Using second order perturbation theory, SFF can be written as\cite{roth},
\begin{equation}
f_s = - \frac{1}{2} \Bra{\psi_0} \hat{\mathcal{T}} \Ket{\psi_0} - \sum\limits_{i \neq 0} \frac{|\Bra{\psi_i} \hat{J_c} \Ket{\psi_0}|^2}{\epsilon
_i - \epsilon_0}, 
\label{fs_0}
\end{equation}
where  $\hat{J_c}=i \sum\limits_l (a_{l+1}^{\dagger}a_{l} - h.c.)$ and $\hat{\mathcal{T}}=-\sum\limits_l (a_{l+1}^{\dagger}a_{l} + h.c.)$ are current operator and usual kinetic energy operator respectively. From the Hamiltonian given in Eq.\ \ref{eq1} one can obtain eigenstates $\Ket{\psi_i}$ with eigenvalue $\epsilon_i$. As shown in Fig.\ \ref{zeroT}(a), the SFF decreases with increasing strength of AA potential $\lambda$ and vanishes at $\lambda=2$, from where IPR of the ground state starts increasing indicating localization transition at $\lambda=2$. From the comparison of SFF and IPR for different values of quasi-periodicity $\alpha$ it is evident that both the quantities decreases for decreasing values of $\alpha$, however the localization transition does not depend on it. 

The variation of IPR with $\lambda$ for all the eigenstates is presented as a surface plot in Fig.\ \ref{zeroT}(b). This shows not only the ground state but all the eigenstates undergo a localization transition at $\lambda=2$ without any mobility edge reflecting the self-dual nature of the AA model.
%
\section{Bosons at finite temperature in the AA-potential}\label{sec3}
In this section we consider non interacting Bose gas in the AA potential at finite temperatures. 
Our main aim is to study the relevant thermodynamic properties of the ideal Bose gas in order to detect the localization phenomena at finite temperature and to understand the the behavior of the Bose gas in different regions of $T-\lambda$ plane. For bosons it is natural to investigate the influence of quasi periodic disorder on formation of Bose-Einstein condensate which is discussed in subsection \ref{sec3.1}.
In subsection \ref{sec3.2} we study the superfluidity and localization of Bose gas at finite temperatures. Finally in subsection \ref{sec3.3} we investigate the signature of condensation and localization phenomena from the entanglement entropy of the gas.
\subsection{\bf Bose-Einstein condensation and crossover temperature}\label{sec3.1}
In ideal Bose gas the macroscopic occupancy of the ground state (single particle lowest energy state) is commonly known as Bose-Einstein condensation. In 3D this quantum phenomena occurs suddenly below a critical temperature showing a phase transition in the thermodynamic limit which can be identified from the singular behavior of certain thermodynamic quantities\cite{pathria}.
As it is well known that the finite temperature phase transition in a one dimensional Bose gas is absent  in the thermodynamic limit \cite{pathria}, we do not expect a `true Bose-Einstein condensation' in this system. However a finite temperature crossover from thermal gas to a quasi-condensate can be observed in finite systems\cite{ketterle,shlyapnikov} due to the presence of ground state energy gap $\Delta_g \sim 1/L$ in a system of size $L$. 
In a weakly interacting 1D Bose gas, phase coherence is developed when the coherence length becomes much larger compared to the typical inter-particle separation leading to the formation of `quasi-condensate' phase which has been studied in details both theoretically and experimentally\cite{castin_rev,shlyapnikov,shlyap_rev,rzazewski_qc,walraven_qc}.
It is important to note that even for 3D Bose gas a sharp transition to BEC is smeared out in the presence of a confined geometries, however the crossover region can be identified from the enhanced fluctuations in certain thermodynamic quantities.
The crossover region of harmonically trapped Bose gas can efficiently be captured from the peak of the ground state number fluctuation\cite{rzazewski_peak,holthaus_peak} and the temperature corresponding to the peak can be identified as BEC crossover temperature which approaches to the critical temperature in the thermodynamic limit\cite{rzazewski_cross}.
The number fluctuation in finite boson system crucially depends on the choice of ensemble since the thermodynamic quantities obtained from different ensemble are not equivalent in general, particularly for finite systems\cite{bec_fluc,weiss}.
Although the thermodynamics of quantum systems are commonly studied within grand canonical ensemble,  
but it turns out that the canonical ensemble is more appropriate to describe such isolated finite system without any exchange of particles with the reservoir
\cite{bec_fluc,weiss,rzazewski_peak,holthaus_peak}.
In what follows, we study the formation of 1D quasi-condensate in AA model and calculate the corresponding crossover temperature from the ground state number fluctuation in both grand canonical and canonical ensemble related to two different physical situation and address the issue of ensemble equivalence.
 
\begin{figure}
 \centering
 \includegraphics[width=5.3cm,height=4.7cm]{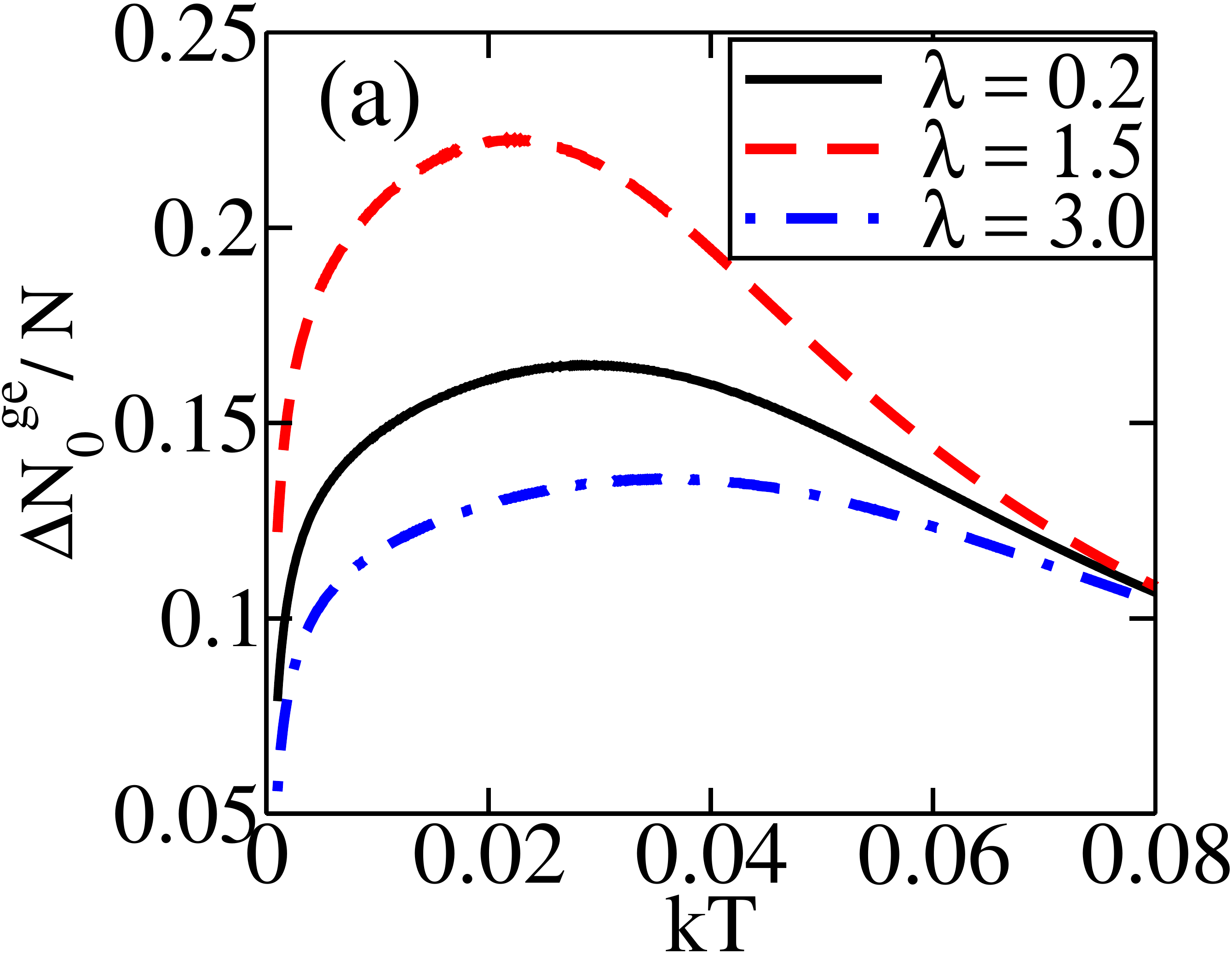}~
 \includegraphics[width=5.3cm,height=4.7cm]{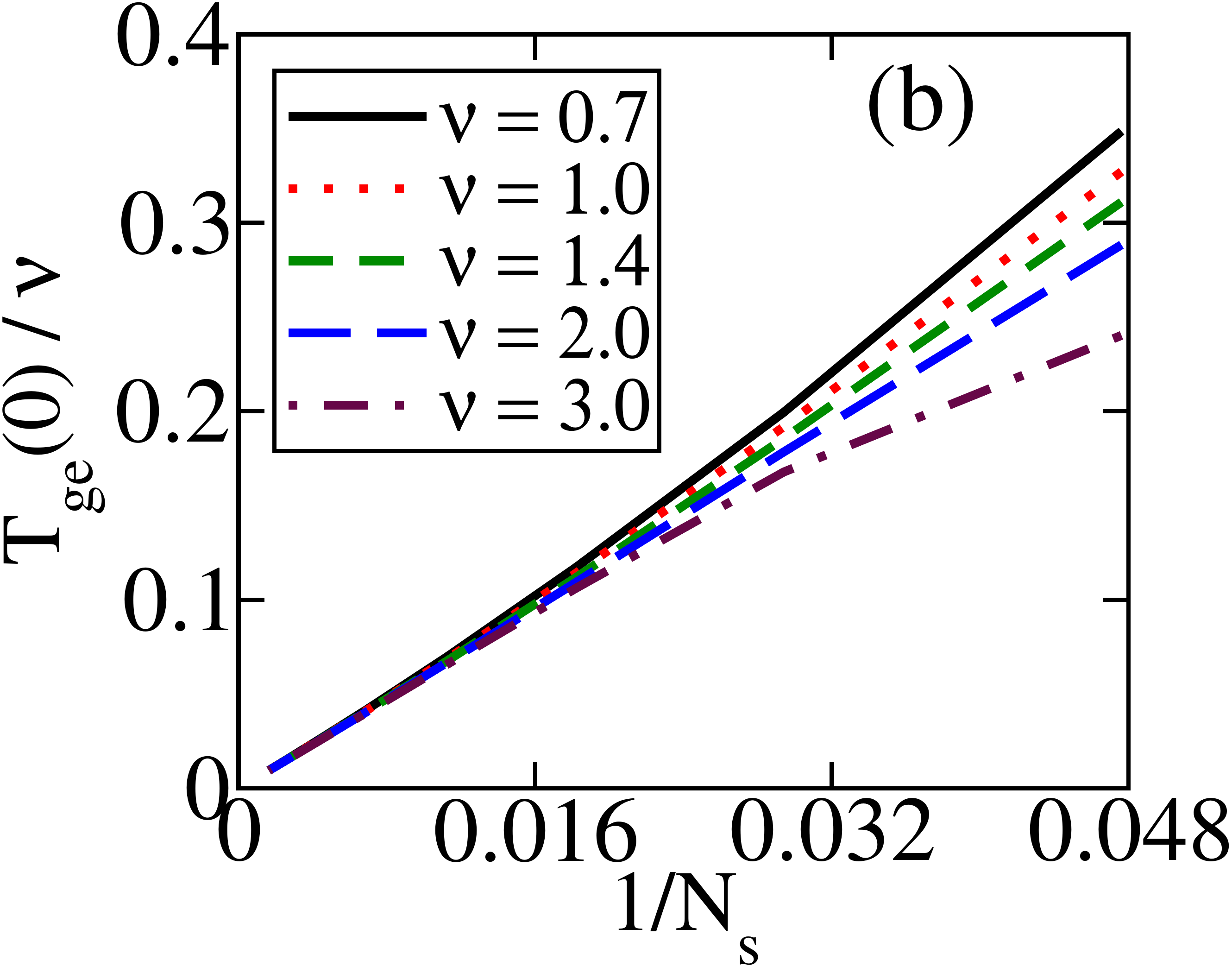}
 \caption{(a) The relative change in ground state occupation $\Delta {N_0}^{ge}/N$ as a function of temperature $kT$(in units of $J$) for different values of $\lambda$. The crossover  temperature $T_{ge}(\lambda)$ is obtained from the maximum of each curve. For this plot $N_s=144$, $\nu=1.0$ and $\alpha=\alpha_g$. (b) Variation of $T_{ge}(0)$(in units of $J$) with number of lattice sites $N_s$ for different values of $\nu$ and $\alpha=\alpha_g$.}
 \label{Nge} 
 \end{figure}

First we discuss the relative change in the ground state occupation, which is defined within the grand canonical ensemble as \cite{blakie},   
\begin{equation}
\Delta {N_0}^{ge}=\frac{1}{{N_0}^{ge}} \bigg|\frac{\partial {N_0}^{ge}}{\partial T}\bigg|
\label{nfluc_gc}
\end{equation}
where,
\begin{equation} 
\frac{1}{{N_0}^{ge}} \bigg|\frac{\partial {N_0}^{ge}}{\partial T}\bigg|=\bigg|(\frac{e^{\beta(\epsilon_0-\mu)}}{e^{\beta(\epsilon_0-\mu)}-1})(\frac{\epsilon_0-\mu}{T^2} + \frac{1}{T} \frac{\partial\mu}{\partial T})\bigg|.
\end{equation}
Here $\epsilon_i$'s are the energy of the single particle states, ${N_0}^{ge}$ is the ground state occupation and $\mu$ is the chemical potential. Above quantity attains a maximum at a certain temperature ($T_{ge}$), by which we can identify the crossover from the condensate phase to the thermal gas. The ground state number fluctuations $\Delta {N_0}^{ge}$ as a function of temperature for different coupling strengths $\lambda$ of the quasi-periodic potential are shown in Fig.\ \ref{Nge}(a).
For different values of $\lambda$ we calculate the crossover temperature $T_{ge}(\lambda)$ from the peak of the number fluctuations $\Delta {N_0}^{ge}$.
This crossover temperature $T_{ge}(\lambda)$ depends on the system size $N_s$, filling of bosons $\nu$ and coupling strength of AA potential $\lambda$. It is important to note that $T_{ge}$ has a strong system size dependence since it vanishes in the thermodynamic limit. The dependence of $T_{ge}(0)$ on $N_s$ in absence of the quasi-periodic potential is shown in Fig.\ \ref{Nge}(b) for different filling.
This behavior of the critical temperature of 1D Bose gas in a lattice of size $L$ can be analyzed from the approximate relation,
\begin{eqnarray}
N/L=2\int_{k_0}^{\pi} \frac{dk}{2\pi} \frac{1}{e^{\beta_{c} J (\cos(k\bar{a}) - 1)} -1} 
\approx2 \int_{k_0}^{k_T} \frac{dk}{\pi \beta_c J (k\bar{a})^2}\nonumber \\
\end{eqnarray}
where, the momentum cutoffs are given by $k_0 = 2\pi/L$ and $\beta_c J (k_{T}\bar{a})^{2} \approx 1$ and $N/L$ is the density of the gas. From this relation we obtain $kT_{c}/J \sim \frac{\pi^{2} \nu}{N_s} + O(1/N_{s}^{3/2})$, where $\nu$ is filling factor of the Bose gas in the lattice of sites $N_s$. A similar scaling of $T_{ge}(0)$ with the lattice sites $N_s$ is revealed in Fig.\ \ref{Nge}(b). Numerically we found that $T_{ge}(0)\sim5.8 \nu/N_s$ for large $N_s$.
\begin{figure}
\centering
\includegraphics[width=5.3cm,height=4.7cm]{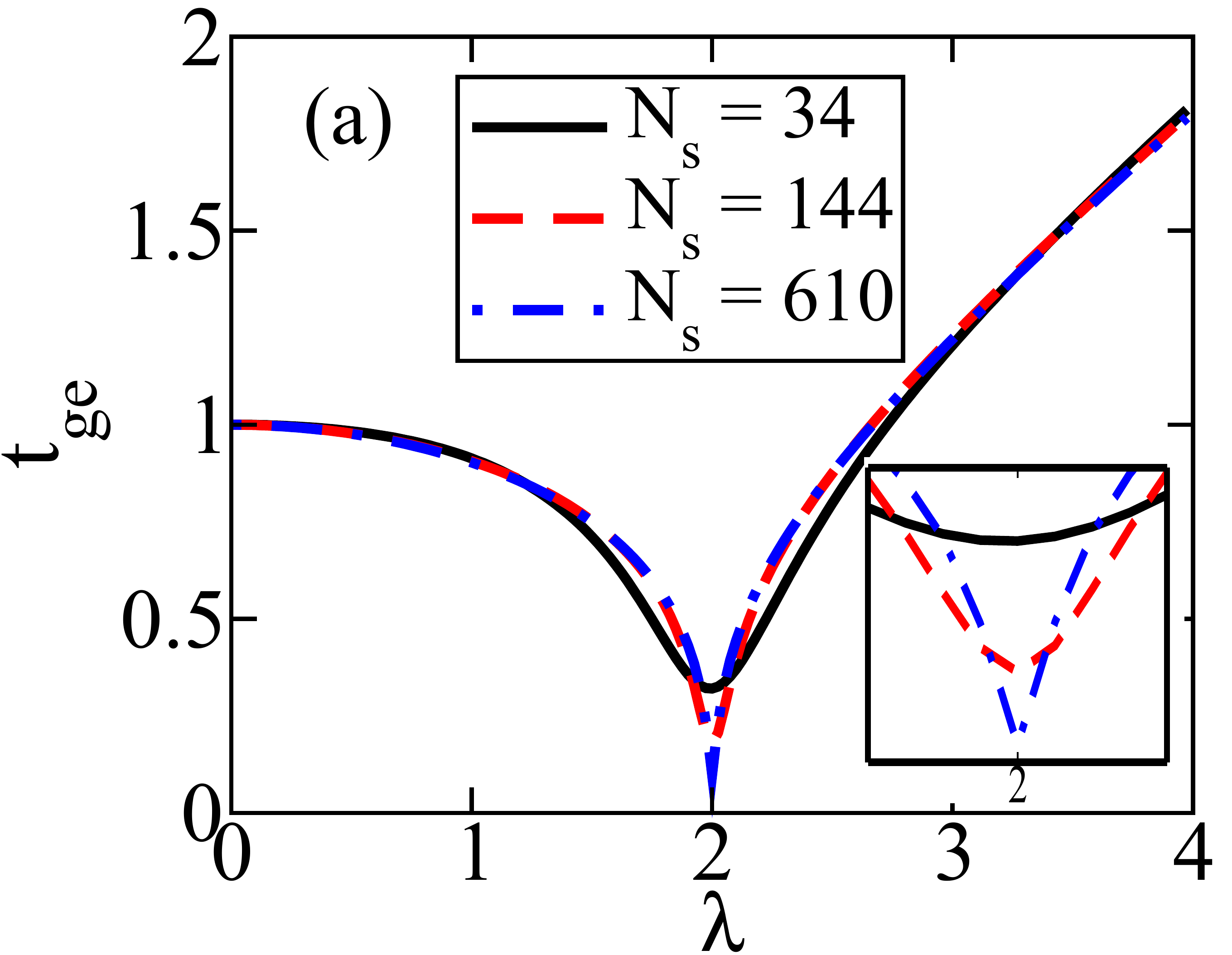}~
 \includegraphics[width=5.3cm,height=4.7cm]{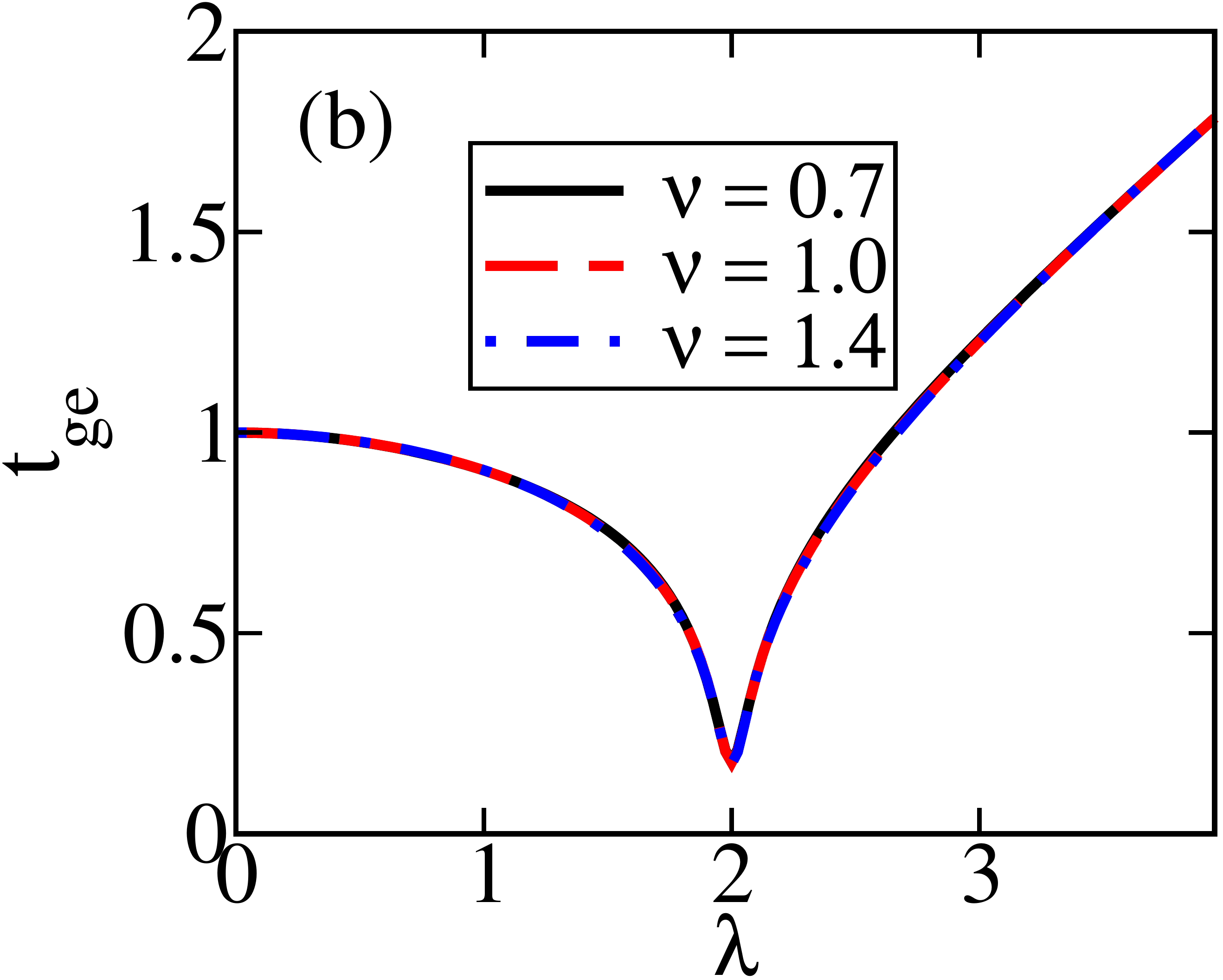}
\caption{Dimensionless crossover temperature $t_{ge}(\lambda)$ as a function of $\lambda$ (a) for increasing values of $N_s$ and $\nu=1.0$; (b) for increasing values of filling $\nu$ and $N_s=144$. In the inset of (a) $t_{ge}(\lambda)$ is zoomed near the critical point $\lambda=2$ showing the finite size effect. For all the plots $\alpha=\alpha_g$.}
\label{tge} 
\end{figure}
In order to analyze the effect of quasi-periodic potential on the crossover temperature corresponding to the Bose-Einstein condensation, we calculate the scaled temperature $t_{ge}(\lambda) = T_{ge}(\lambda)/T_{ge}(0)$ as a function of coupling strength $\lambda$. For bosons with a given filling fraction $\nu$ the scaled crossover temperature $t_{ge}(\lambda)$ as a function of $\lambda$ is shown in Fig.\ \ref{tge}(a) for different lattice size $N_s$. {As seen from Fig.\ \ref{tge}(a), the scaling of crossover temperature eliminates the finite size effect and reveals following interesting features arising due to the AA potential.
In the delocalized regime ($\lambda <2$) the stability of the quasi-condensate phase under thermal fluctuation is reduced due to the presence of quasi-periodic disorder resulting in a decrease of the crossover temperature $T_{ge}$ with increasing $\lambda$. At the critical coupling $\lambda_c =2, $ $T_{ge}$ becomes vanishingly small. However, for $\lambda >2$ the crossover temperature $T_{ge}$ increases with the coupling strength $\lambda$.   
As seen from the inset of Fig.\ \ref{tge}(a), at the critical point $T_{ge}(\lambda_c)$ approaches to zero for increasing system size. This is an interesting manifestation of the self-dual critical point of AA-model related to the localization transition.
 At the critical point the effect of large quantum fluctuation can destroy the condensate even at zero temperature leading to the vanishing of $T_{ge}$ in the thermodynamic limit. This behavior of the crossover temperature is typically observed in quantum critical phenomena.
In this context it is also interesting to point out that for large system size the maximum of number fluctuation $\Delta N^{ge}$ at $\lambda =2$ diverges as $\sim N_{s}^{1.4}$ as a result of quantum fluctuation arising from vanishing of energy gap at the critical point. 
In Fig.\ \ref{tge}(b), the dimensionless quantity $t_{ge}(\lambda)$ is shown as a function of $\lambda$ for different values of filling $\nu$.
From Fig.\ \ref{tge}(a) and (b) a universal feature 
of the scaled crossover temperature $t_{ge}(\lambda)$ emerges which shows that for sufficiently large system size the variation of $t_{ge}$ with $\lambda$ is almost independent of filling $\nu$ and lattice size $N_s$ except in the close vicinity of the critical point $\lambda_{c}=2$. Although the absolute value of the crossover temperature is non-universal and vanishes for large system size, a suitable scaling enables to capture the universal features of BEC crossover on the quasi-periodic potential strength.

We now consider the thermodynamics of fixed number of bosons within canonical ensemble which is found out to be more appropriate description of finite quantum systems\cite{bec_fluc,weiss,rzazewski_peak,holthaus_peak}. 
In isolated quantum systems like ultracold atoms in a trap, the particle exchange with the heat bath can be neglected.
To investigate the crossover phenomena corresponding to BEC in 1D quasi-periodic potential we calculate the ground state number fluctuations.     
In order to obtain the thermodynamic quantities we first calculate the partition function using the following recursion relation \cite{borrmann,weiss,tran}.
\begin{equation}
Z_N (\beta) = \frac{1}{N}\sum\limits_{j=1}^N (\pm 1)^{j+1} Z_1 (j\beta) Z_{N-j} (\beta),
\label{partition}
\end{equation}
where $Z_{N}(\beta)$ is the partition function of $N$ particles at a temperature $kT = 1/\beta$ and `$+$'(`$-$') stands for bosons(fermions). We will assume $k = 1$ hereafter in this paper. 
In the canonical ensemble, the probability distribution of the occupancy of the $i$th energy state is given by\cite{weiss},
\begin{equation}
P_i(n_i)=e^{-n_i \beta \epsilon_i} \frac{Z_{N-n_i}}{Z_N} - e^{-(n_i + 1) \beta \epsilon_0} \frac{Z_{N-n_i-1}}{Z_N},
\label{Pn0}
\end{equation}
From the probability $P_0(n_0)$ the occupancy of the ground state is given by,
\begin{equation}
{N_0}^{ce}=\sum\limits_{n_0=0}^{N} n_0 P_0(n_0).
\label{can_occu}
\end{equation}
Similarly the fluctuation of the ground state occupancy can be calculated using \cite{weiss,tran},
\begin{equation}
\Delta {N_0}^{ce} = \sqrt{{{{N_0}^{sq}}}-{({N_0}^{ce}})^2},
\label{can_fluc} 
\end{equation}
where ${{{N_0}^{sq}}}=\sum\limits_{n_0=0}^{N} {n_0}^2 P_0(n_0)$.

From above mentioned thermodynamic quantities the crossover to the condensate phase can be identified. As shown in Fig.\ \ref{Nce}(a), near the crossover the occupancy of first excited state shows a peak and becomes comparable with the ground state population. This indicates an appreciable number of particles are thermally excited from the condensate leading to a crossover to thermal gas phase.
Also the maximum of the probability distribution of ground state occupancy changes from zero to a finite number for decreasing temperature indicating the formation of condensate with large occupancy of the ground state (see Fig.\ \ref{Nce}(b)). As discussed earlier the peak in the ground state number fluctuation depicted in Fig.\ \ref{Nce}(c) signifies the crossover to BEC and corresponding crossover temperature $T_{ce}$ can be obtained from the position of the peak\cite{rzazewski_cross}. 
From the finite size analysis of the crossover temperature for $\lambda =0$ (shown in Fig.\ \ref{Nce}(b)), we find $T_{ce}(0)\sim 8.8\nu/N_s$ for large $N_s$ showing the similar scaling with $N_s$ and $\nu$ as obtained earlier for grand canonical ensemble. The temperature $T_{ce}(\lambda)$ obtained from the canonical ensemble shows similar behavior as $T_{ge}(\lambda)$ with the coupling strength $\lambda$ of the AA potential and vanishes at the critical coupling $\lambda_c = 2$. However, the crossover temperature obtained in the number conserving canonical ensemble is larger than that obtained from the grand canonical ensemble. As discussed earlier, the scaling of the crossover temperature $t_{ce} = T_{ce}(\lambda)/T_{ce}(0)$ shows similar universal dependence on $\lambda$ without any appreciable effects of $\nu$ and $N_s$.
\begin{figure}
\centering
\includegraphics[width=5.2cm,height=4.7cm]{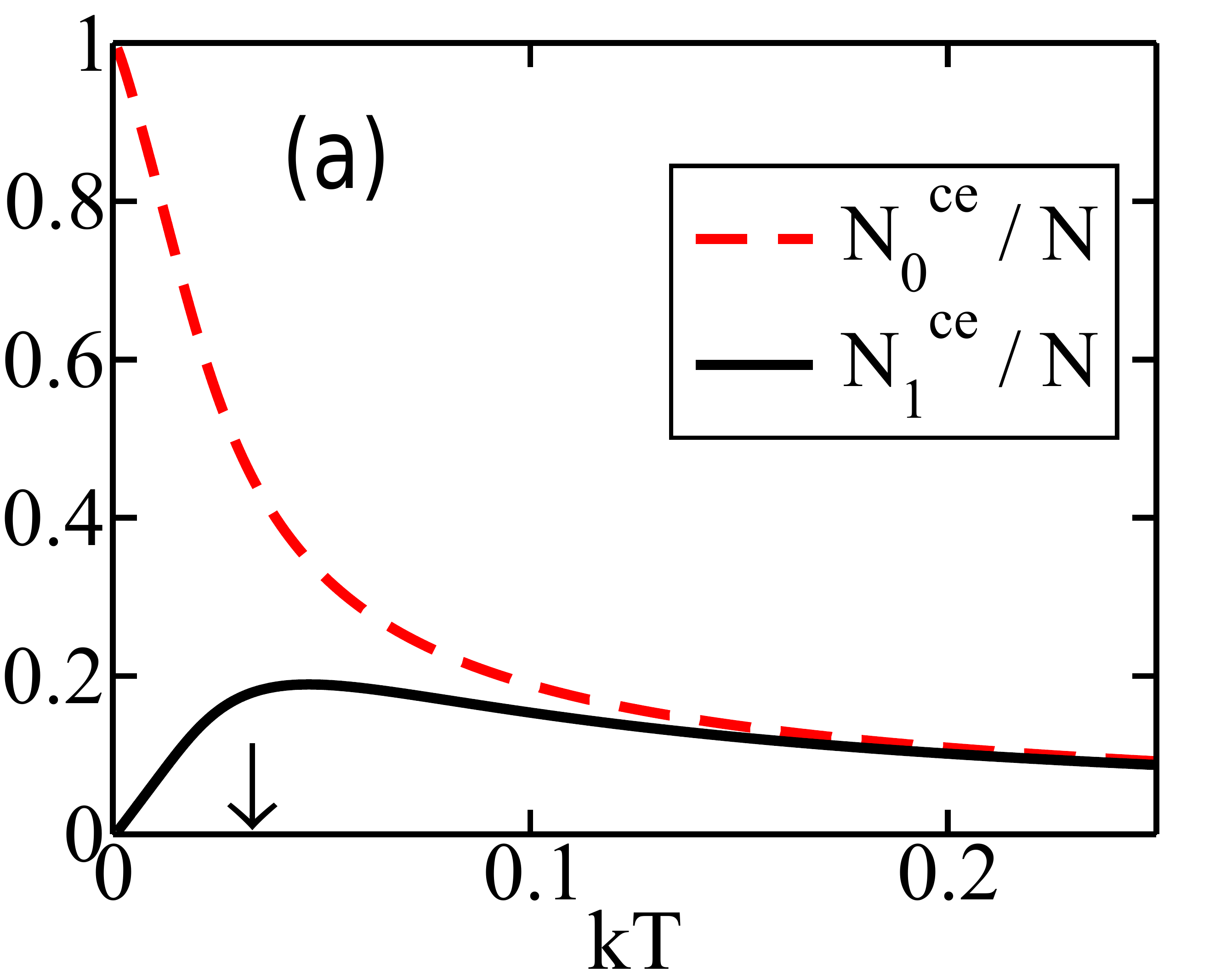}~
 \includegraphics[width=5.2cm,height=4.7cm]{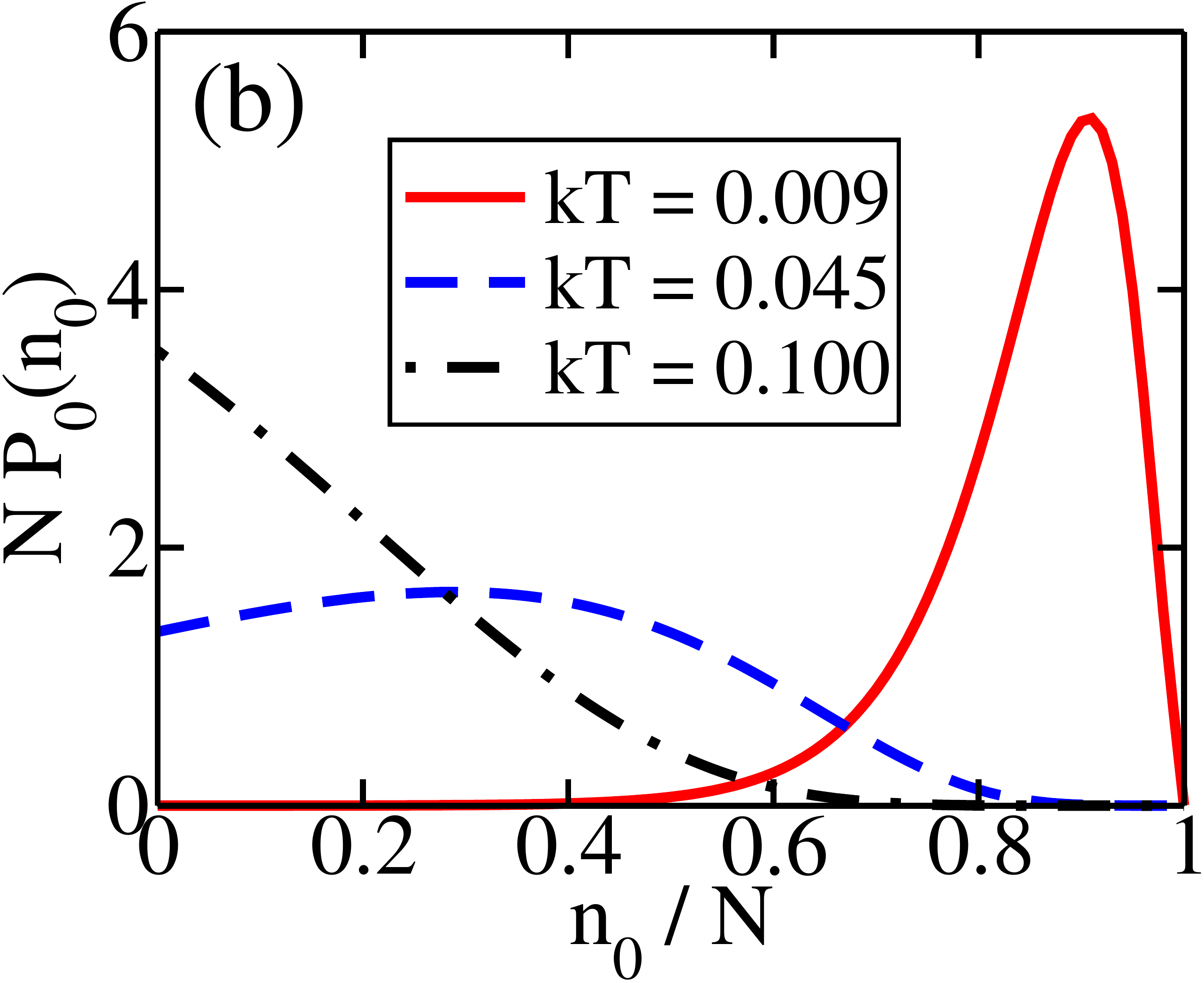}
\includegraphics[width=5.7cm,height=4.6cm]{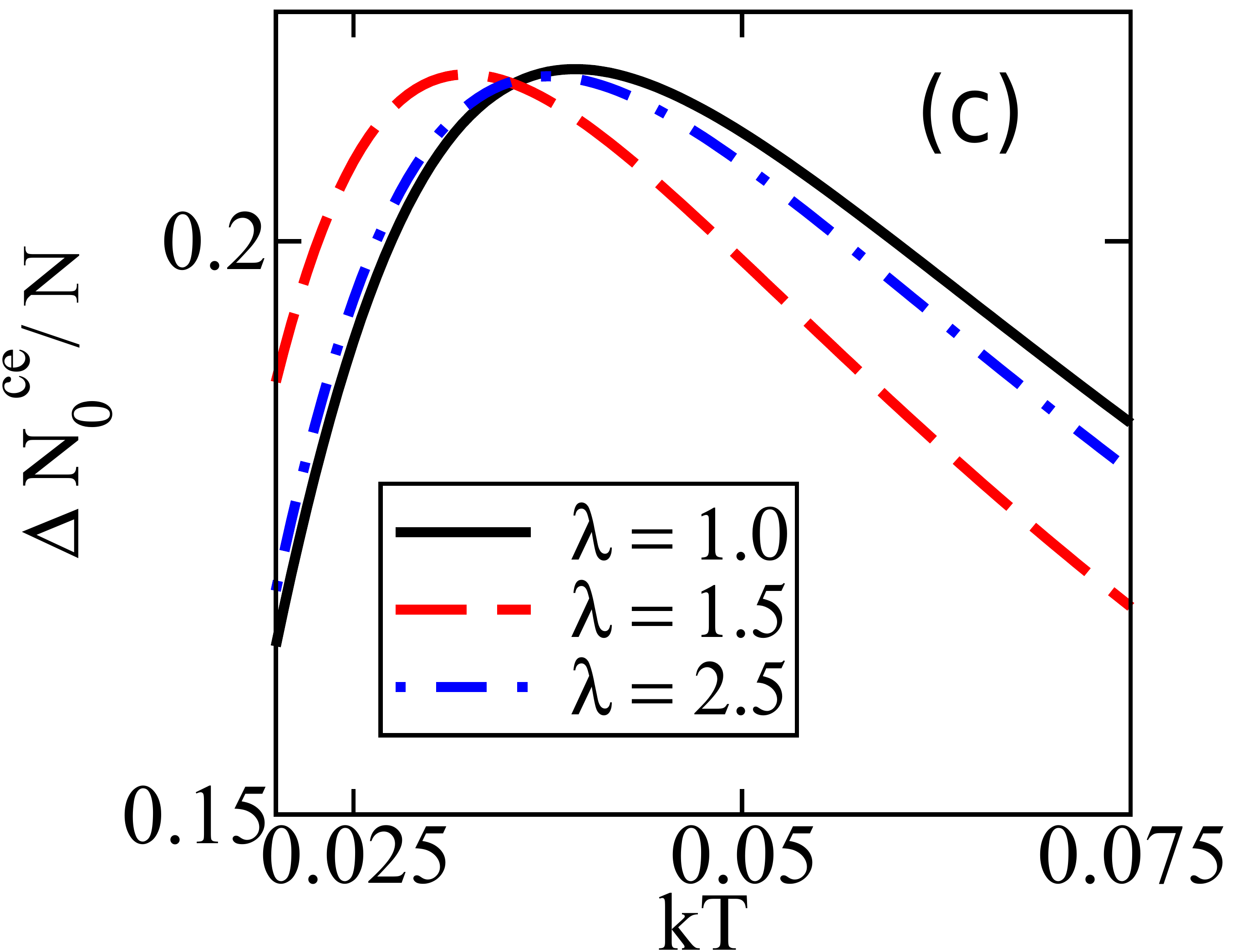}~
 \includegraphics[width=5.5cm,height=4.7cm]{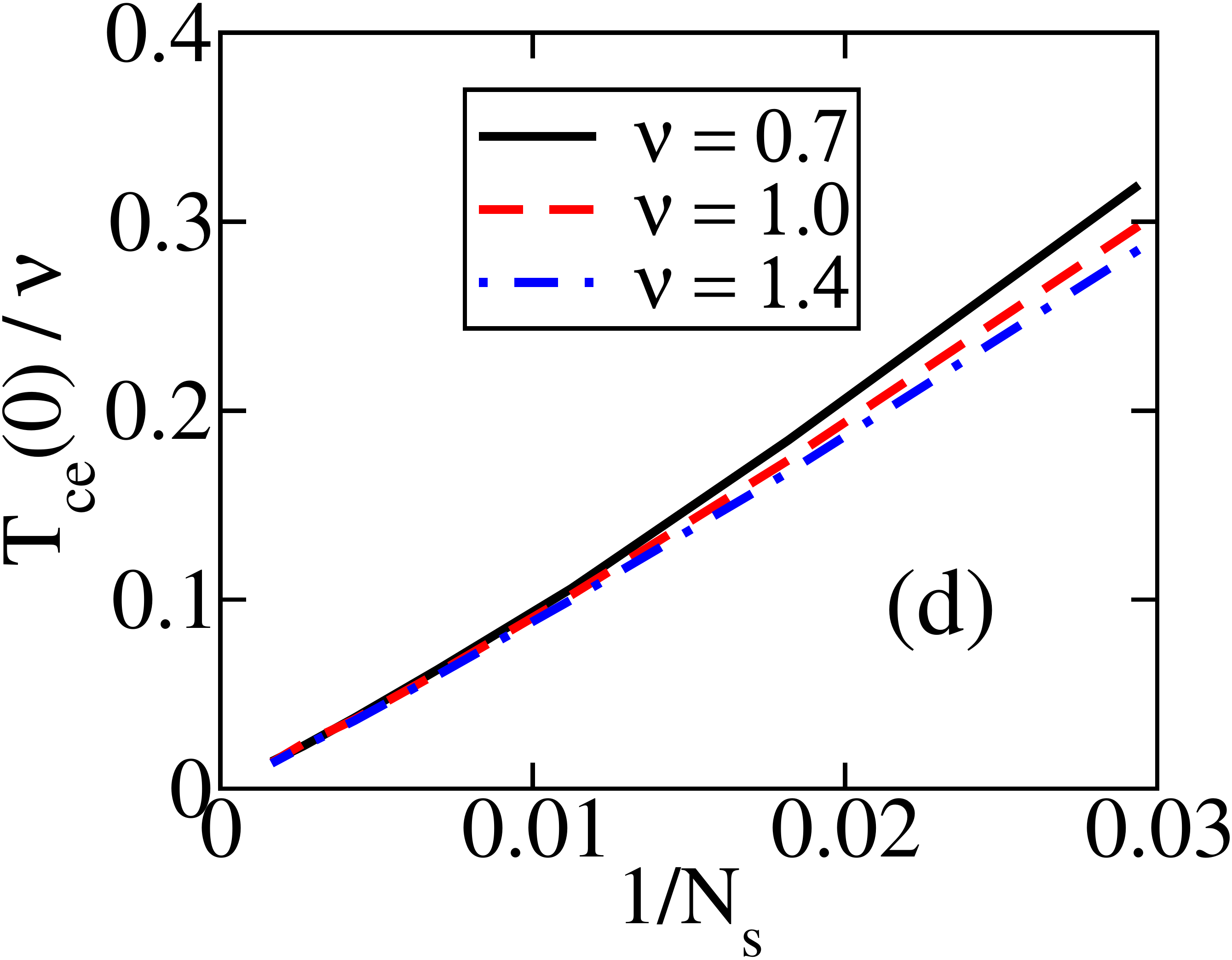}
\caption{(a) The occupancy of the ground state ${N_0}^{ce}$ and the first excited state  ${N_1}^{ce}$ as a function of temperature $kT$(in units of $J$) for $\lambda=1.5$. The down-arrow indicates the temperature $T_{ce}$. (b) The probability distribution of the ground state occupancy $P_0(n_0)$ corresponding to $\lambda=1.5$. (c) The ground state number fluctuation ${\Delta N_0}^{ce}/N$ as a function of temperature $kT$(in units of $J$) for increasing $\lambda$. For figures (a-c) filling $\nu=0.7$ and $N_s=144$. (d) Dependence of $T_{ce}(0)$(in units of $J$) on $N_s$ for different filling $\nu$. For all plots $\alpha=\alpha_g$}
\label{Nce}
\end{figure}

An interesting feature of the crossover phenomena is captured as the
dimensionless scaled crossover temperatures $t_{ge}(\lambda)$ and $t_{ce}(\lambda)$ corresponding to two different ensembles are compared with the scaled energy gap $\Delta_0=\Delta(\lambda)/\Delta(0)$ as depicted in Fig.\ \ref{gap}, where $\Delta(\lambda)$ is the ground state energy gap at the coupling strength $\lambda$. 
An equivalence between these scaled dimensionless quantities is evident from Fig.\ \ref{gap}, which can be written as,
\begin{equation}
t_{ge}(\lambda) \simeq t_{ce}(\lambda) \simeq \Delta_{0}(\lambda),
\label{equivalence}
\end{equation}
for different values of the quasi-periodicity $\alpha$ of the AA model. 
From the above analysis following conclusions can be drawn:\\
i) functional dependence of the scaled crossover temperature on coupling strength $\lambda$ is related to the scaled energy gap of the ground state of the AA model which  is independent of the filling $\nu$ of bosons and has finite size effect only in the close vicinity of the critical coupling for sufficiently large system size .\\
ii) the scaled crossover temperatures $t_{ge/ce}(\lambda)$ has monotonic dependence on quasi-periodicity $\alpha$ of the AA model.\\
iii) `ensemble equivalence' between dimensionless scaled crossover temperatures $t_{ge}$ and $t_{ce}$ obtained from grand canonical and canonical ensembles, although the absolute value of the crossover temperatures $T_{ge/ce}$ differ. 
\begin{figure}
\centering
\includegraphics[width=9.0cm,height=7.0cm]{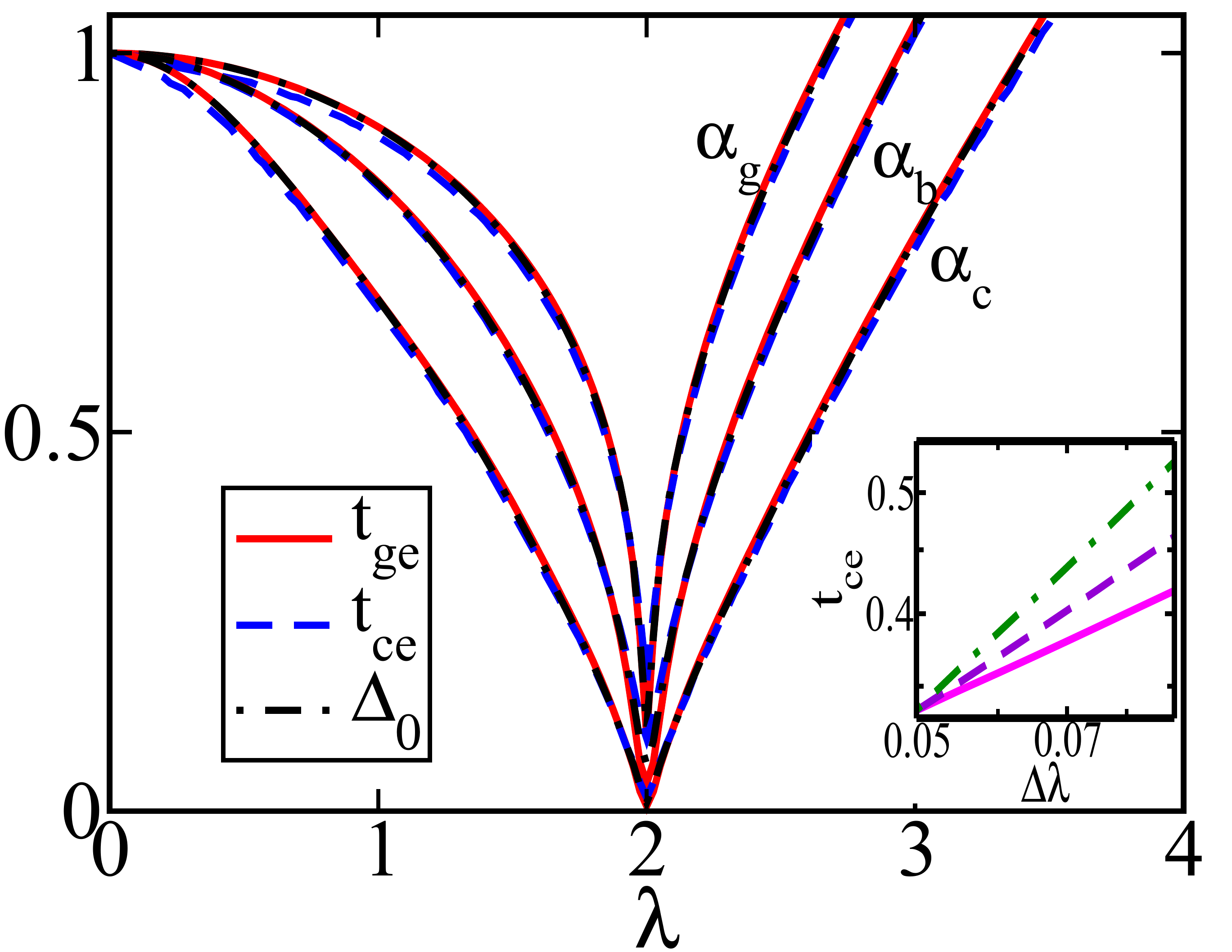}
\caption{Universal dependence of the scaled crossover temperatures $t_{ge}$, $t_{ce}$ and the scaled ground state energy gap $\Delta_0$ on $\lambda$ for decreasing values of $\alpha$. The inset shows the variation of $t_{ce}$, (properly scaled for clarity) with $\Delta \lambda =|\lambda-\lambda_c|$ in the log-log scale for $\alpha=\alpha_g$(solid), $\alpha_b$(dashed) and $\alpha_c$(dot dashed). 
Here $\nu=1.0$ and $N_s=610$ for $\alpha=\alpha_g$.}
\label{gap}
\end{figure}

The decrease of the scaled crossover temperature $t_{ge/ce}$ with decreasing quasi-periodicity $\alpha$ indicates that the formation of the quasi-condensate phase is more stable under thermal fluctuations for higher values of $\alpha$. However, irrespective of the value of quasi-periodicity $\alpha$ the crossover temperature always vanishes at the self-dual critical point $\lambda_c = 2$ following power law $t_{ce}\sim{|\lambda-\lambda_c|}^{\gamma(\alpha)}$ as a result of quantum fluctuation associated with vanishing of energy gap $\Delta(\lambda)$.    
Close to $\lambda_c$, the variation of $t_{ce}$ with $| \lambda -\lambda_c |$ in log-log scale is depicted in the inset of Fig.\ \ref{gap} which shows that the exponent of the power law $\gamma(\alpha)$ decreases with increasing $\alpha$.
Numerically we find that $\gamma(\alpha) = 0.38, 0.56$
and $0.80$ for $\alpha_g$ , $\alpha_b$ and $\alpha_c$ respectively.
Vanishing of the energy gap and the critical temperature at the critical point following power law are typical features of quantum critical systems.     
The dependence of crossover temperature on both coupling strength and quasi-periodicity can be tested in cold atom experiments by changing relative frequency and intensity of the bichromatic optical lattice. Enhanced fluctuations near the transition point can be probed by the momentum distribution of the atom cloud.
\subsection{\bf Superfluidity and localization phenomena at finite temperature}\label{sec3.2}
To investigate the localization of bosons at finite temperature in the presence of a quasi-periodic potential we calculate the superfluid fraction (SFF) and IPR at non zero temperatures. A natural extension of the definition of SFF at finite temperature is given by\cite{fisher}, 
\begin{equation}
{f_s}^\beta=\frac{{N_s}^2}{N} \frac{F(\theta) - F(0)}{{\theta}^2}, 
\label{sff_t}
\end{equation}
where $F$ stands for the Helmholtz free energy and $\theta$ is the phase twist. 
From the Hamiltonian given in Eq.\ \ref{ham_twist} and computing the second derivative of the free energy a closed form expression for SFF at finite temperature is obtained in \cite{giamarchi} which matches with the Eq.\ \ref{fs_0} at zero temperature.
At low temperatures the ground state dominates and contribution from other 
states is suppressed by a factor $e^{-\beta \Delta}$, where $\Delta$ is the energy gap between the ground state and the first excited state. Next we calculate the SFF of the Bose gas at non zero temperatures using Eq.\ \ref{sff_t}. 
Both temperature and quasi-periodic potential destroy superfluidity as shown in Fig\ \ref{SFF_T}(a). For a given $\lambda$, SFF vanishes at a temperature ${{T_1}^*(\lambda)}$ which can be denoted as the superfluid crossover temperature. 
In absence of any disorder, similar to $T_{ge}(0)$,the superfluid crossover temperature $T_{1}^*(0)$ also becomes proportional to $\nu/N_s$ for large system size which is shown in Fig.\ \ref{SFF_T}(b). From our numerical calculation we find ${T_1}^*(0)\sim 14.5\nu/N_s$ . The superfluid crossover temperature $T_{1}^*(\lambda)$ is larger than the quasi-condensation crossover temperature $T_{ge}(\lambda)$.
\begin{figure}
\centering
\includegraphics[width=5.6cm,height=4.7cm]{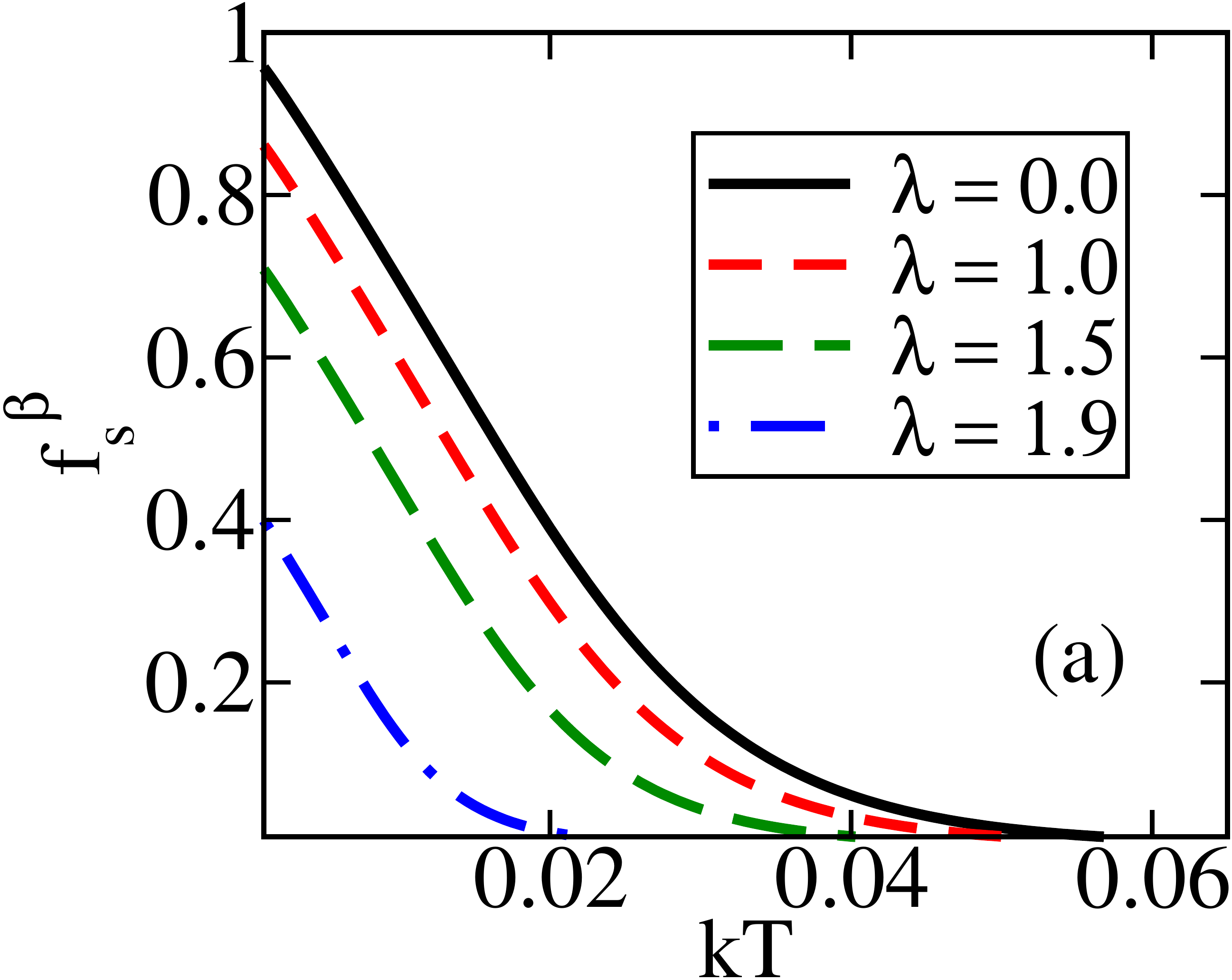}~
\includegraphics[width=5.6cm,height=4.7cm]{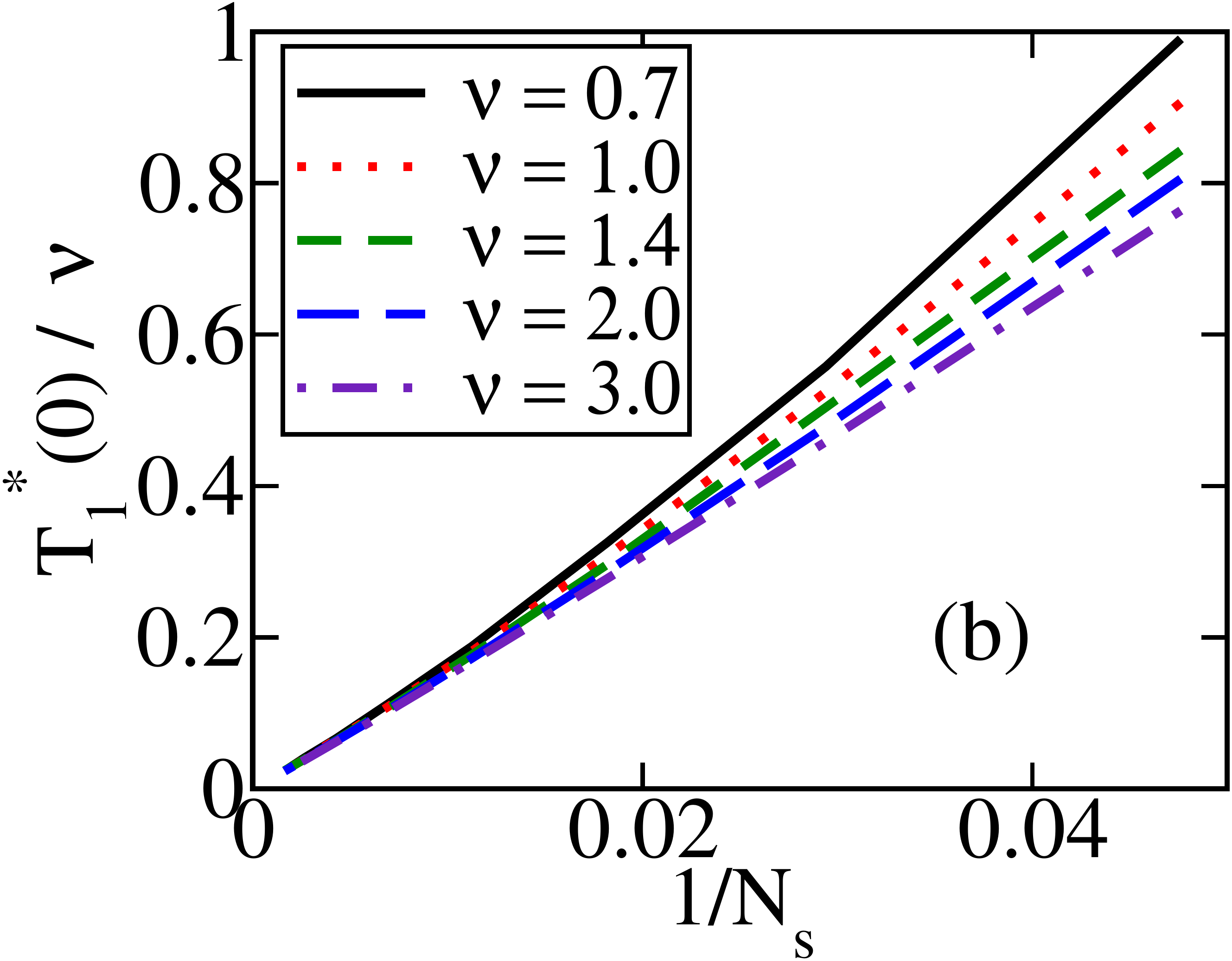}
\caption{(a) The SFF ${f_s}^\beta$ as function of temperature $kT$ (in units of $J$) for increasing $\lambda$. At $T_{1}^{*}(\lambda)$ ${f_s}^\beta$ vanishes. For this plot $\nu=0.7$ and $N_s=144$. (b) Dependence of the superfluid crossover temperature ${T_1}^*(0)$ (in units of $J$) with $\lambda =0$ on $N_s$ for different filling $\nu$. For all plots $\alpha=\alpha_g.$}
\label{SFF_T}
\end{figure}
Similar to $t_{ge}$, we define scaled crossover temperature corresponding to the superfluid phase as $t_{1}^*=T_{1}^*(\lambda)/T_{1}^*(0)$ which is plotted as a function of $\lambda$ in Fig.\ \ref{t1_star}. This dimensionless crossover temperature $t_{1}^*$ decreases with $\lambda$ and vanishes at the critical coupling strength $\lambda=2$ for sufficiently large $N_s$. However, for small system size the SFF vanishes at $\lambda > 2$ as seen from Fig.\ \ref{t1_star}(a). The finite size effects and filling fraction $\nu$ dependence of $t_{1}^*(\lambda)$ are depicted in Fig.\ \ref{t1_star}(a). It is evident from these results that the dimensionless quantity $t_{1}^*(\lambda)$ corresponding to the superfluid crossover becomes independent of $\nu$ and a universal feature emerges for large $N_s$, although the crossover temperature $T_{1}^*(\lambda)$ remains non-universal and vanishes in the thermodynamic limit. 
Similar to BEC crossover temperature, the scaled superfluid crossover temperature $t_{1}^{*}(\lambda)$ decreases with decreasing quasi-periodicity $\alpha$, however they vanish for $\lambda > 2$ due to the absence of superfluidity in the localized regime (see Fig.\ \ref{t1_star}(b)). 
\begin{figure}
\centering
\includegraphics[width=5.3cm,height=4.7cm]{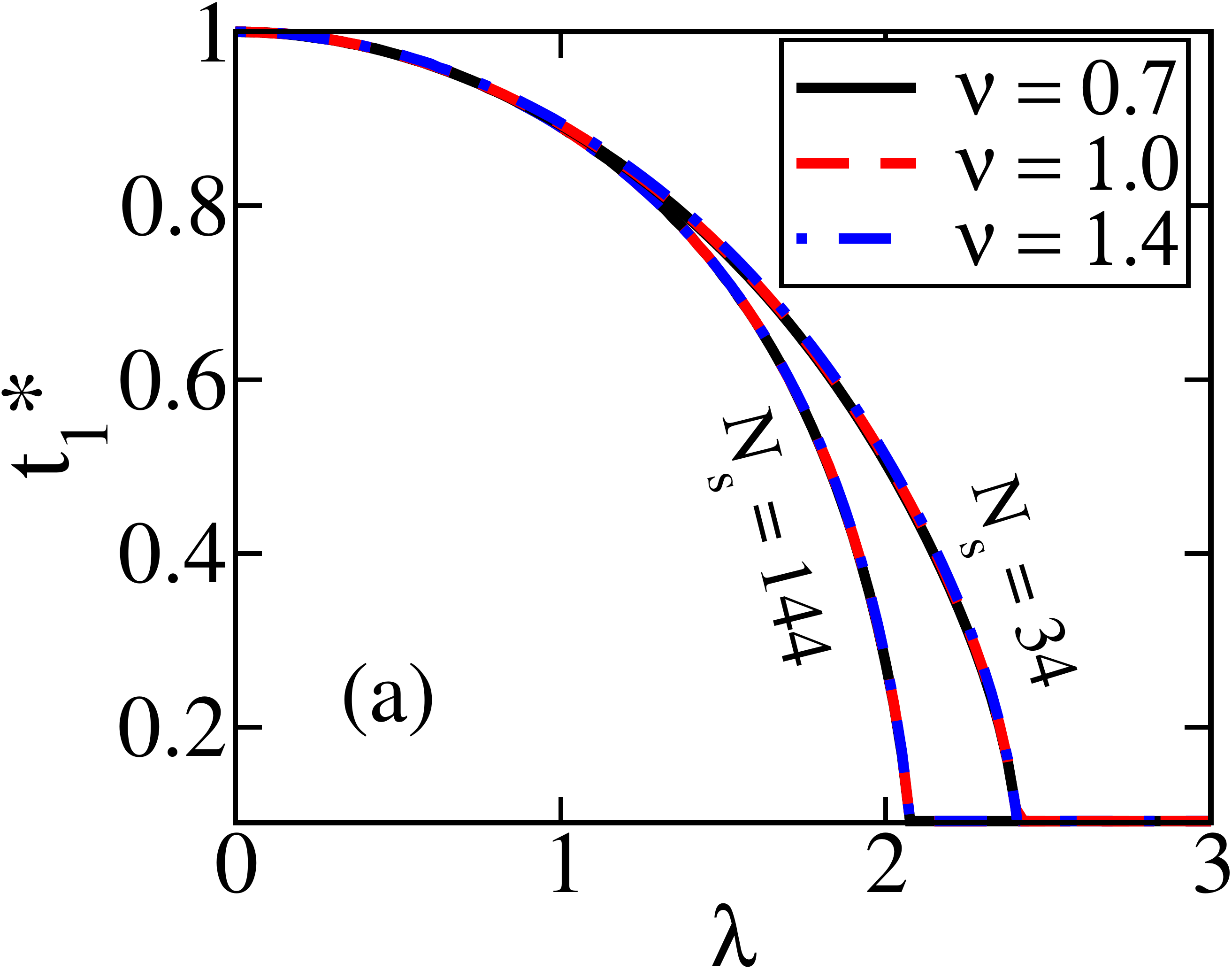}~
 \includegraphics[width=5.3cm,height=4.7cm]{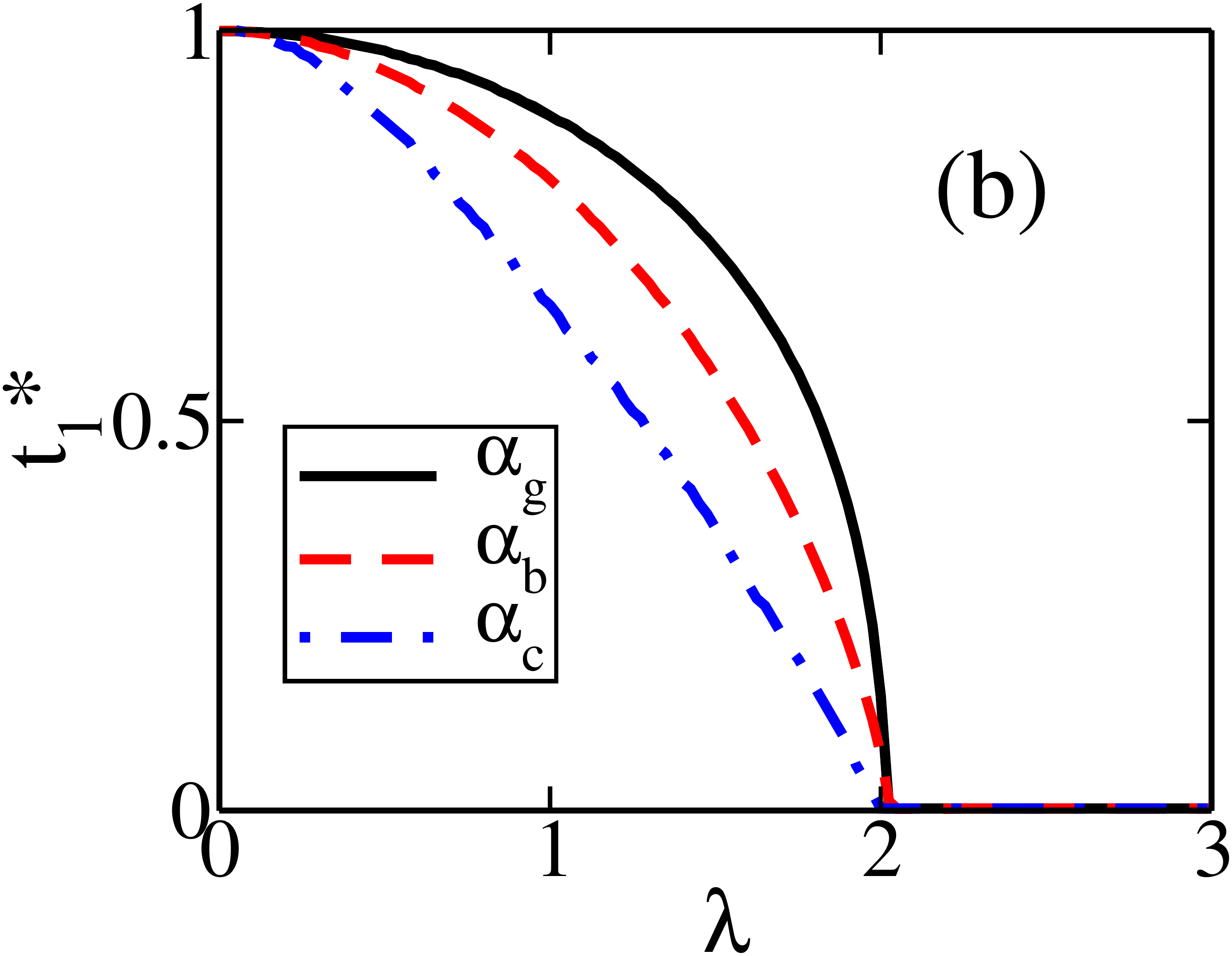} 
\caption{(a) Variation of the dimensionless superfluid crossover temperature ${t_1}^*$ with $\lambda$ for increasing filling $\nu$ and two fixed values of $N_s=34, 144$ and $\alpha=\alpha_g$. (b) The scaled crossover temperature ${t_1}^*$ as function of $\lambda$ for decreasing $\alpha$. For this plot $\nu=1.0$ and $N_s=610$ for $\alpha=\alpha_g$.}
\label{t1_star}
\end{figure}

Next we investigate the localization of the Bose gas from its density distribution.
Also we define the IPR at finite temperature to understand the finite temperature localization in the region $\lambda >2$. 
The finite temperature IPR can be defined as,
\begin{equation} 
I_\beta=\sum\limits_l  {\rho_n(l)}^2 ,
\end{equation}
where $\rho(l) = \sum\limits_i \frac{{|\phi_i (l)|}^2}{e^{\beta(\epsilon_i-\mu)}-1}$ is the density at site $l$ and $\rho_n (l) = \frac{\rho (l)}{\sum\limits_l \rho (l)}$ is the normalized density. Here $\epsilon_i$ and $\phi_i(l)$ are the $i$th energy level of the single particle spectrum and the amplitude of the corresponding normalized eigenfunction at site $l$ respectively. The IPR scales as $1/N_s$ for fully delocalized system and asymptotically reaches to $1$ for extremely localized system. The IPR $I_{\beta}$ as a function of temperature 
is shown in Fig.\ \ref{ipr_T}(a) for different values of $\lambda$.
From the decay of $I_{\beta}$ it is clear that the thermal fluctuations favor delocalization of the Bose gas.
In Fig.\ \ref{ipr_T}(b) a surface plot of IPR as a function of temperature and the strength of AA-potential $\lambda$ is shown to identify the crossover from localized to thermally disordered Bose gas. A contour with $I_{\beta}=0.5$ is shown by solid line that roughly indicates the change from single-site localization to double-site localization, helping us to understand a localization-delocalization crossover phenomena driven by thermal fluctuations. Corresponding to $I_{\beta}=0.5$ a temperature is denoted by $T_{2}^{*}(\lambda)$. This can also be attributed to the change in the behavior of the correlation length of interacting bosons as a function of temperature \cite{nessi}.
To study the effect of quasi-periodicity $\alpha$ on localization phenomena, we calculate the same contour corresponding to $I_{\beta}=0.5$ for different metallic mean values of $\alpha$. In Fig.\ \ref{EE_boson}(d) the contours are shown in $T/T_{ge}(0) - \lambda$ plane where we use dimensionless temperature $T/T_{ge}(0)$ for better comparison. It is clear that the degree of localization decreases with decreasing values of $\alpha$ and less thermal fluctuation is required for delocalization from single site. This is in agreement with the variation of ground state IPR with $\alpha$ shown in Fig.\ \ref{zeroT}(a).
It is interesting to note that the shape of the contours with equal IPR resembles that of the crossover temperature corresponding to the condensate phase in the regime $\lambda >2$. Existence of the condensate phase with localization for $\lambda>2$ indicates formation of glassy phase in the presence of interactions. 
\begin{figure}
\centering
\includegraphics[width=5.1cm,height=4.7cm]{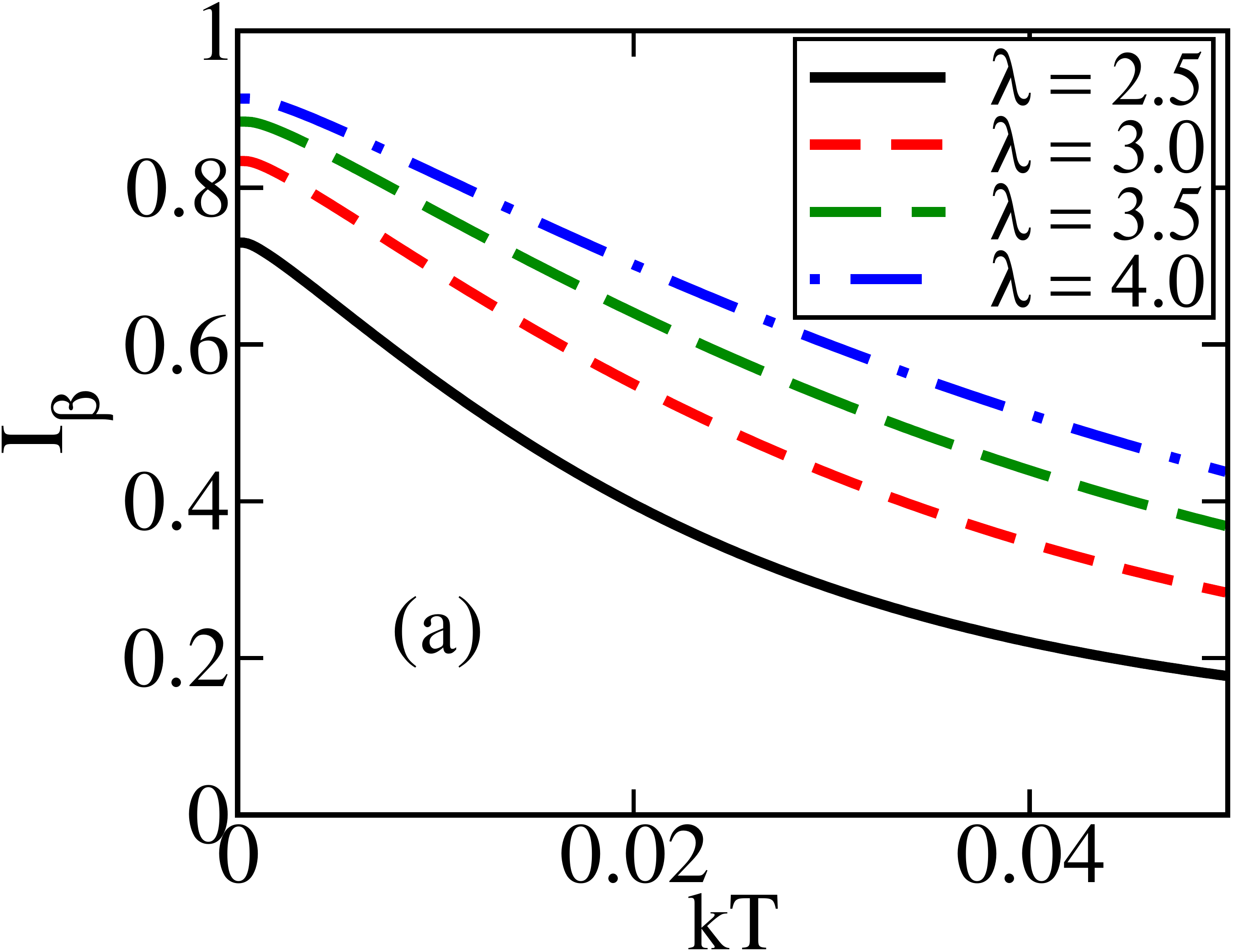}~
 \includegraphics[width=5.7cm,height=5.3cm]{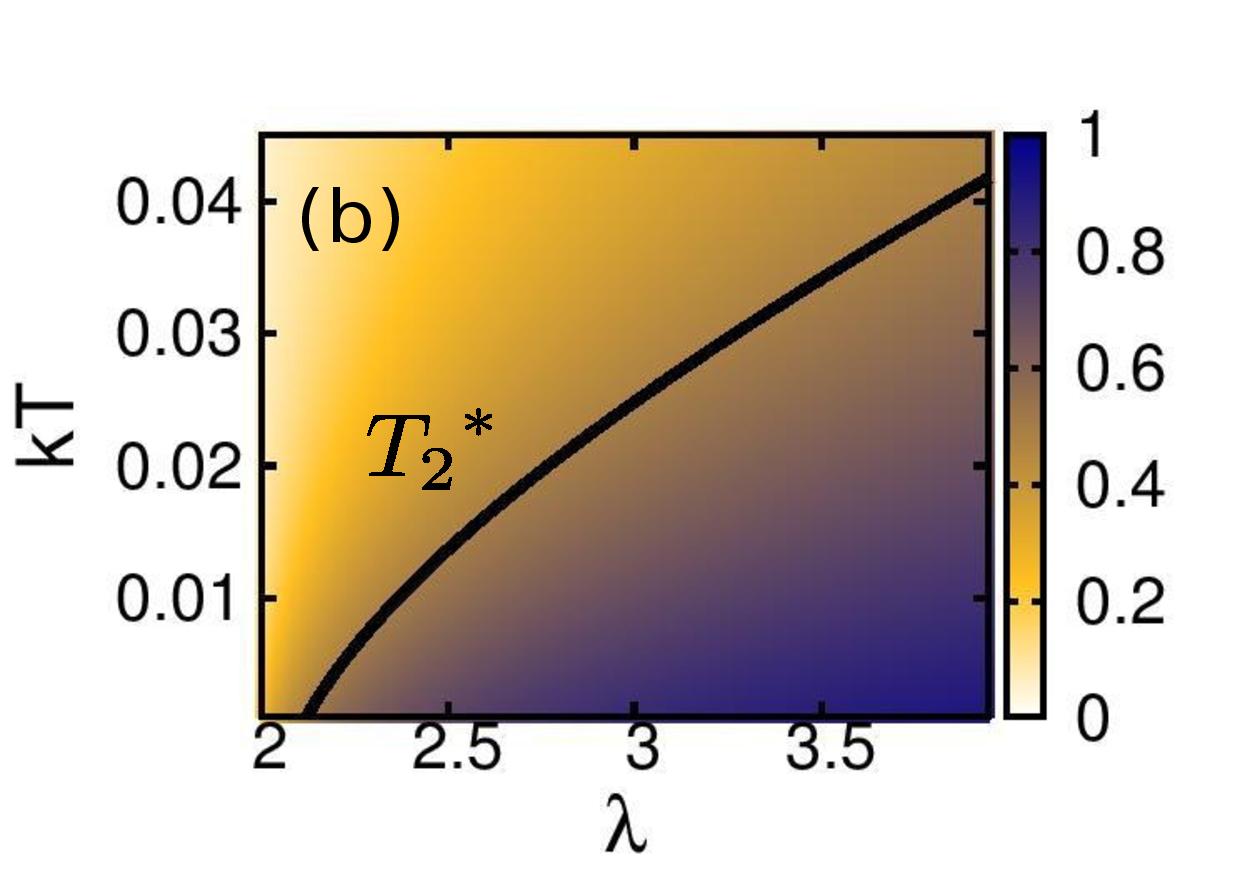}
\caption{(a) Decay of the finite-temperature IPR $I_\beta$ with temperature $kT$ (in units of $J$) for increasing $\lambda$. (b) Surface plot of IPR as a function of $kT$ (in units of $J$) and $\lambda$. A contour with IPR$=0.5$ is shown that denotes single-site to double-site localization transition (${T_2}^*$). For all the plots $\nu=0.7$, $N_s=144$ and $\alpha=\alpha_g$.}
\label{ipr_T}
\end{figure}
\subsection{\bf Entanglement entropy}\label{sec3.3}
To gain a better insight of the localization phenomena and the effect of the critical point $\lambda =2$ at finite temperature, we calculate the single particle entanglement entropy (EE) of the AA model. We divide the full lattice in two equal parts `A' and `B' and calculate the reduced density matrix $\rho_A$ corresponding to the part `A' by tracing out the full density matrix with respect to the basis states of part `B',
\begin{eqnarray}
\rho_{A} = Tr_{B} \rho
\label{red_dm}
\end{eqnarray}
where, $\rho$ is the density matrix(DM) of the total system. To calculate the EE of the non interacting 
bosons (and fermions) one can construct the entanglement Hamiltonian from the correlation matrix \cite{peschel}. However, for bosons at zero temperature the EE diverges as $log(N)$ and does not capture the localization phenomena of the ground state. To capture the localization transition of the single particle wavefunctions we follow a much simpler method to obtain the single particle EE \cite{fradkin}. At zero temperature the full DM is given by $ \rho = |\psi_G\rangle \langle \psi_G|$, where the ground state $|\psi_G\rangle $ can be written as,
\begin{equation}
|\psi_G\rangle = \sum_{l \in A} \phi_{0} (l) a_{l}^{\dagger}|0\rangle_{A}\otimes |0\rangle_{B} + \sum_{l \in B} \phi_{0}(l)|0\rangle_{A} \otimes a_{l}^{\dagger} |0\rangle_{B},
\label{gsp}
\end{equation}
where $\phi_{0}(l)$ is the amplitude and $a_{l}^{\dagger}$ is the bosonic creation operator at site $l$; $|0\rangle_{A/B}$ are the vacuum states corresponding to `A' and `B' subsystems respectively.
The matrix elements of the reduced DM corresponding to the single particle basis states of subsystem `A' are given by,
\begin{equation}
\rho_{A}(l,l^\prime) = \phi_{0}^{*}(l) \phi_{0}(l^\prime); \rho_{A}(0,0) = 1 - \sum_{l\in A} |\phi_{0}(l)|^2  
\label{rho_sp}
\end{equation}
where, $\rho_{A}(0,0)$ is the matrix element with respect to the vacuum state $|0\rangle_{A}$. As shown in \cite{fradkin}, this single particle DM has two non vanishing eigenvalues $\lambda_{1} = \sum_{l\in A} |\phi_{0}(l)|^2$, and
$\lambda_{2} = 1 -  \sum_{l\in A} |\phi_{0}(l)|^2$. Hence the single particle EE of the ground state of the AA model is given by,
\begin{equation}
S_{A} = - \lambda_{1}ln\lambda_{1} - \lambda_{2}ln\lambda_{2}.
\label{gsEE}
\end{equation}
For completely delocalized plane wave state the EE takes a maximum value $S_{A} = \ln{2}$ and for single site localized state $S_{A} = 0$. The variation of $S_{A}$ with the potential strength $\lambda$ is shown in Fig.\ \ref{EE_boson}(a), which clearly captures the localization transition at $\lambda_c = 2$. The EE decreases as the wavefunction becomes more localized for higher values of $\lambda$. In the localized regime $\lambda >2$, the EE increases for lower values of quasi-periodicity $\alpha$ indicating lower degree of localization, as observed from the IPR of ground state shown in Fig.\ \ref{zeroT}(a).

Next we generalize this single particle DM at finite temperature, which can be written as,
\begin{equation}
\rho = \sum_{n} p_{n}|\psi_n\rangle \langle \psi_n|
\label{DMT}
\end{equation}
where, $|\psi_{n}\rangle$ is the $n$th eigenstate of the AA model and $p_{n}$ describes the occupation probability at this energy state. For a single particle at temperature $T$, $p_{n} \sim e^{-\epsilon_{n}/kT}$. To incorporate the effect of Bose statistics we consider $p_{n} = \frac{1}{N} \frac{1}{e^{\beta (\epsilon_{n} - \mu)} -1}$; where $\mu$ is the chemical potential for $N$ bosons at temperature $T$. By tracing out the single particle states of the subsystem `B', we obtain the matrix elements of the reduced DM $\rho_{A}(l,l^\prime) = \sum_{n} p_{n} \phi_{n}^{*}(l) \phi_{n}(l^\prime)$, and $\rho_{A}(0,0)= 1 - \sum_{l=1}^{L/2} \rho_{A}(l,l)$. From the eigenvalues of $\rho_{A}$ we calculate the single particle EE at finite temperature which is depicted in Fig.\ \ref{EE_boson}(b) as a contour plot in the $\lambda - T$ plane. Formation of superfluid phase (I), localized phase (II) and disordered thermal Bose gas (III) in different regions of $\lambda- T$ plane are schematically demarcated by the isoentropic contour (represented by solid line in Fig.\ \ref{EE_boson}(b)) resembling the the shape of the crossover temperature $T_{ge/ce}$.
An increase of EE is associated with the superfluid to thermal Bose gas crossover for $\lambda < 2$, and localized to thermally disordered gas in the regime $\lambda > 2$.
A finite temperature manifestation of the critical point $\lambda_{c} = 2$ is evident from Fig.\ \ref{EE_boson}(b).  
In Fig.\ \ref{EE_boson}(c) and (d) the isoentropic contours and contours with fixed IPR are shown in $T/T_{ge}(0) -\lambda$ plane for different values of $\alpha$. For decreasing values of quasi-periodicity $\alpha$ same amount of EE can be generated for lower temperature, which is consistent with observed decrease in crossover temperature as well as IPR with decreasing $\alpha$.
\begin{figure}
\centering
\includegraphics[width=5.0cm,height=4.6cm]{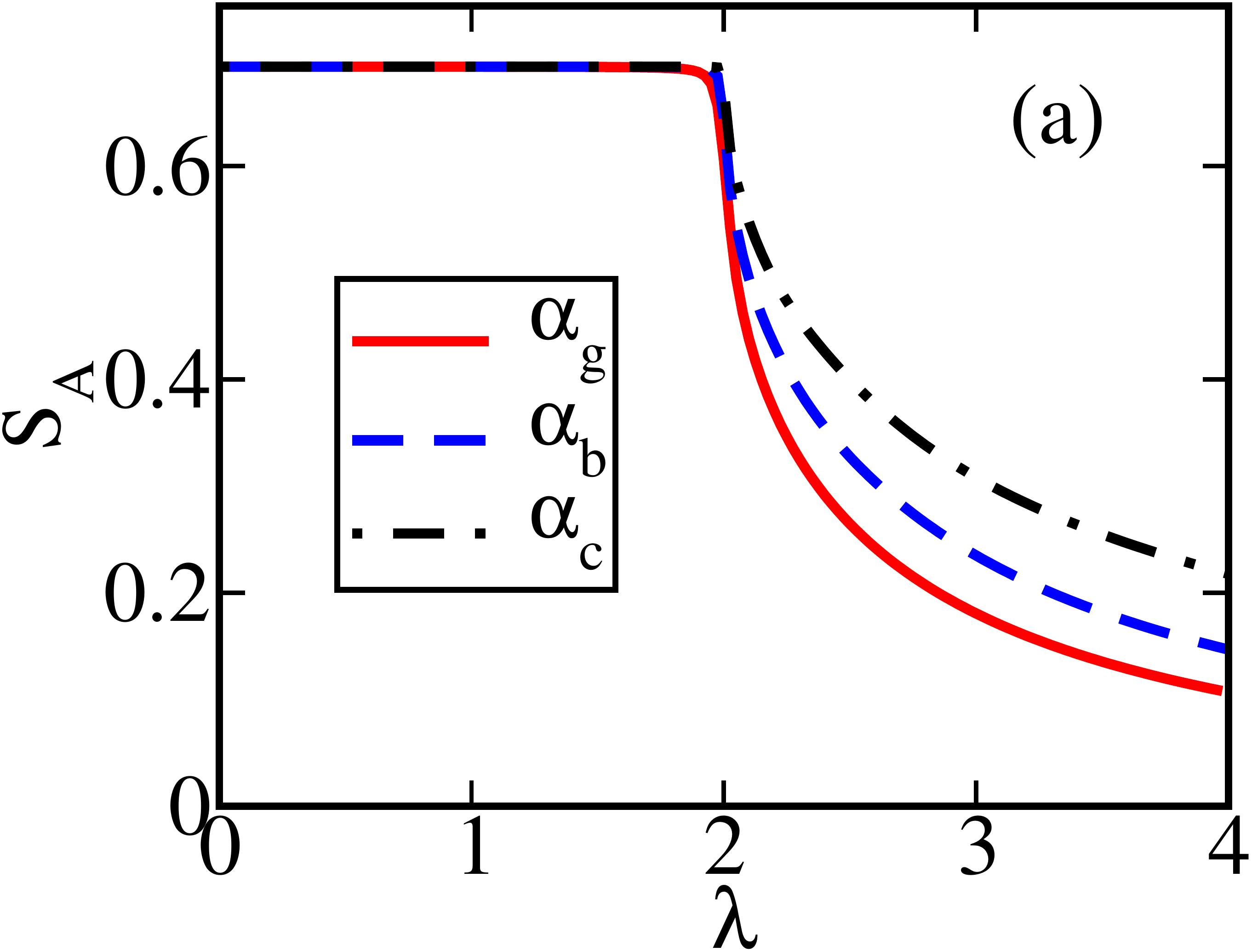}~
 \includegraphics[width=5.7cm,height=5.3cm]{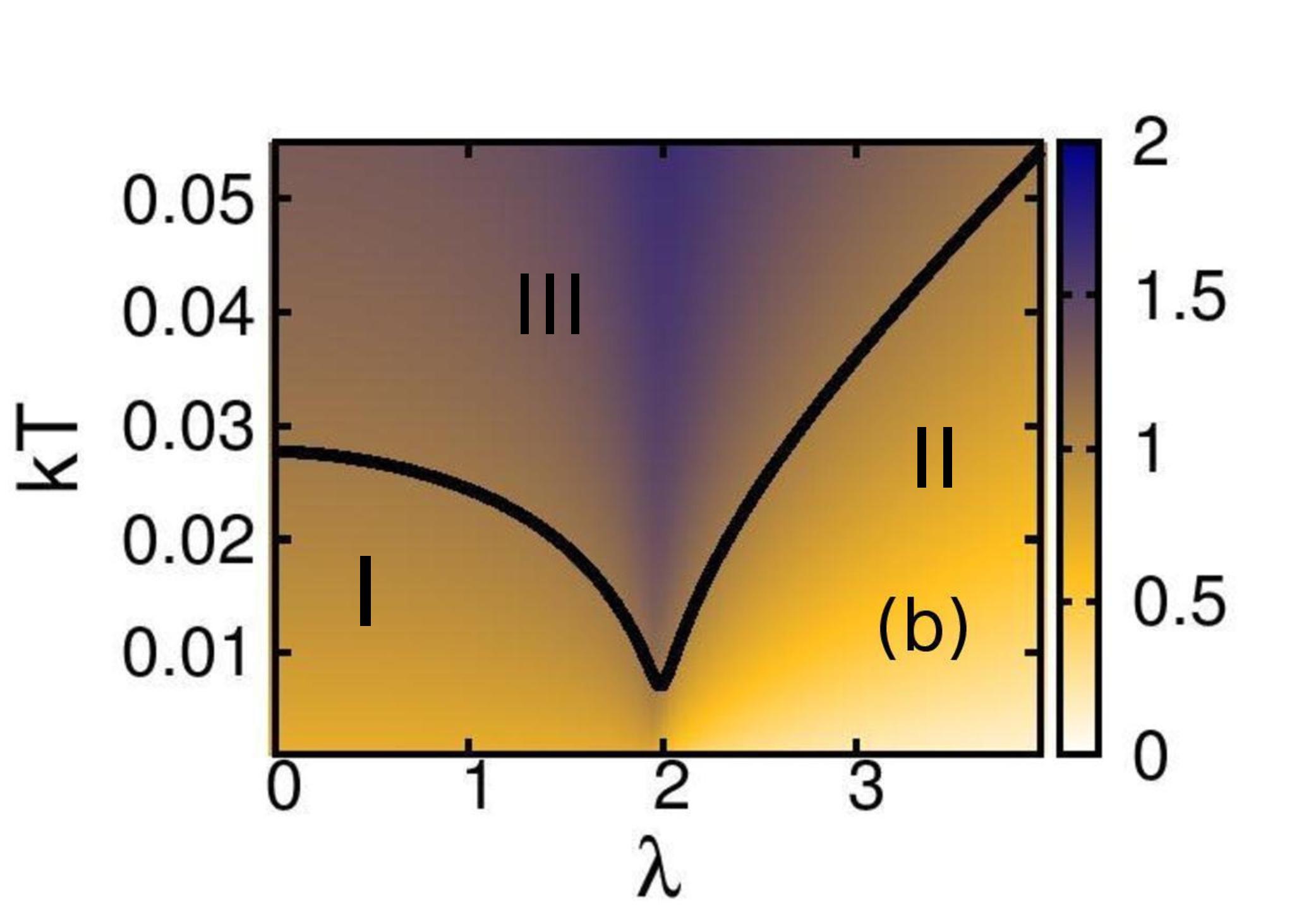}
\includegraphics[width=5.0cm,height=4.6cm]{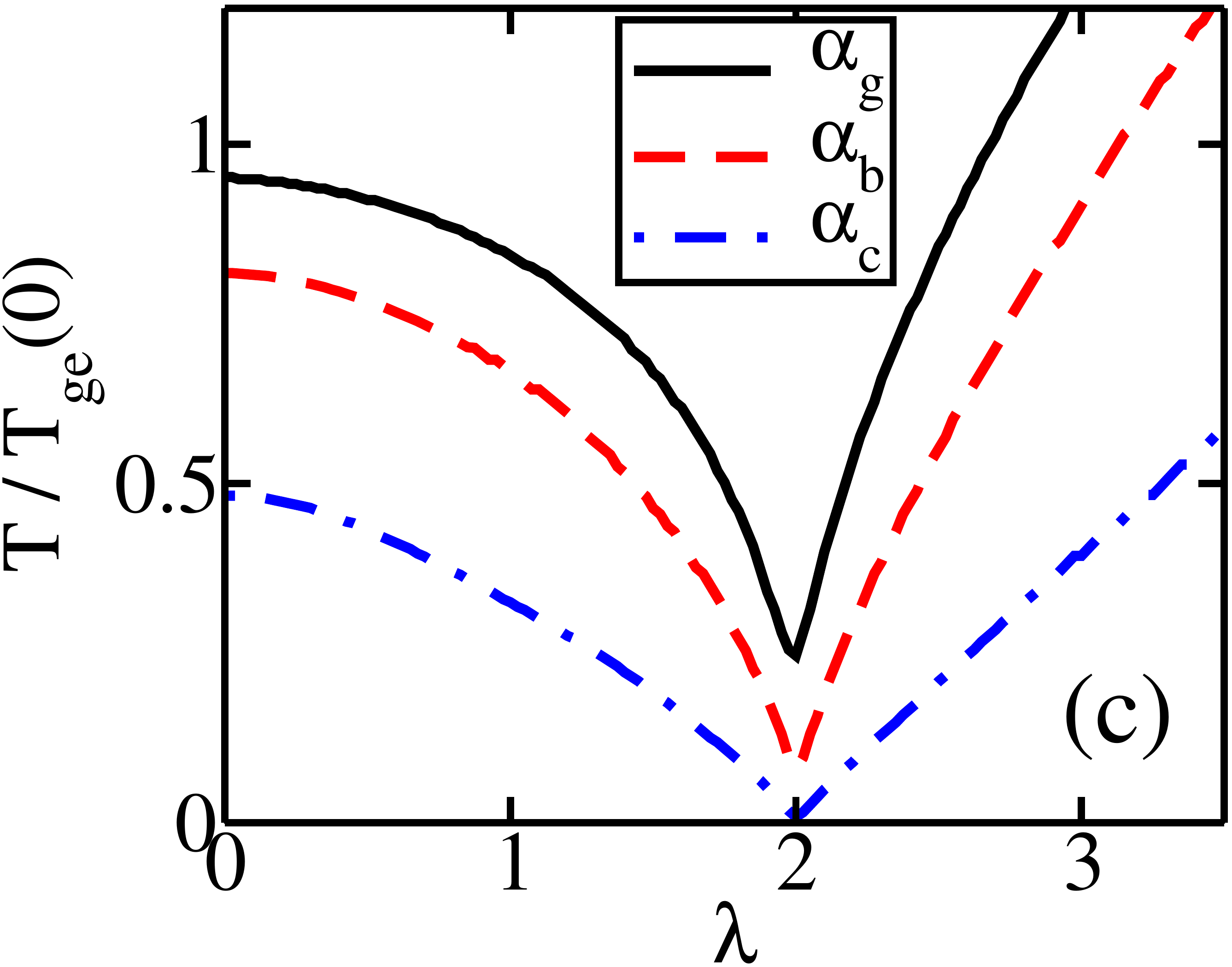}~
\includegraphics[width=5.0cm,height=4.6cm]{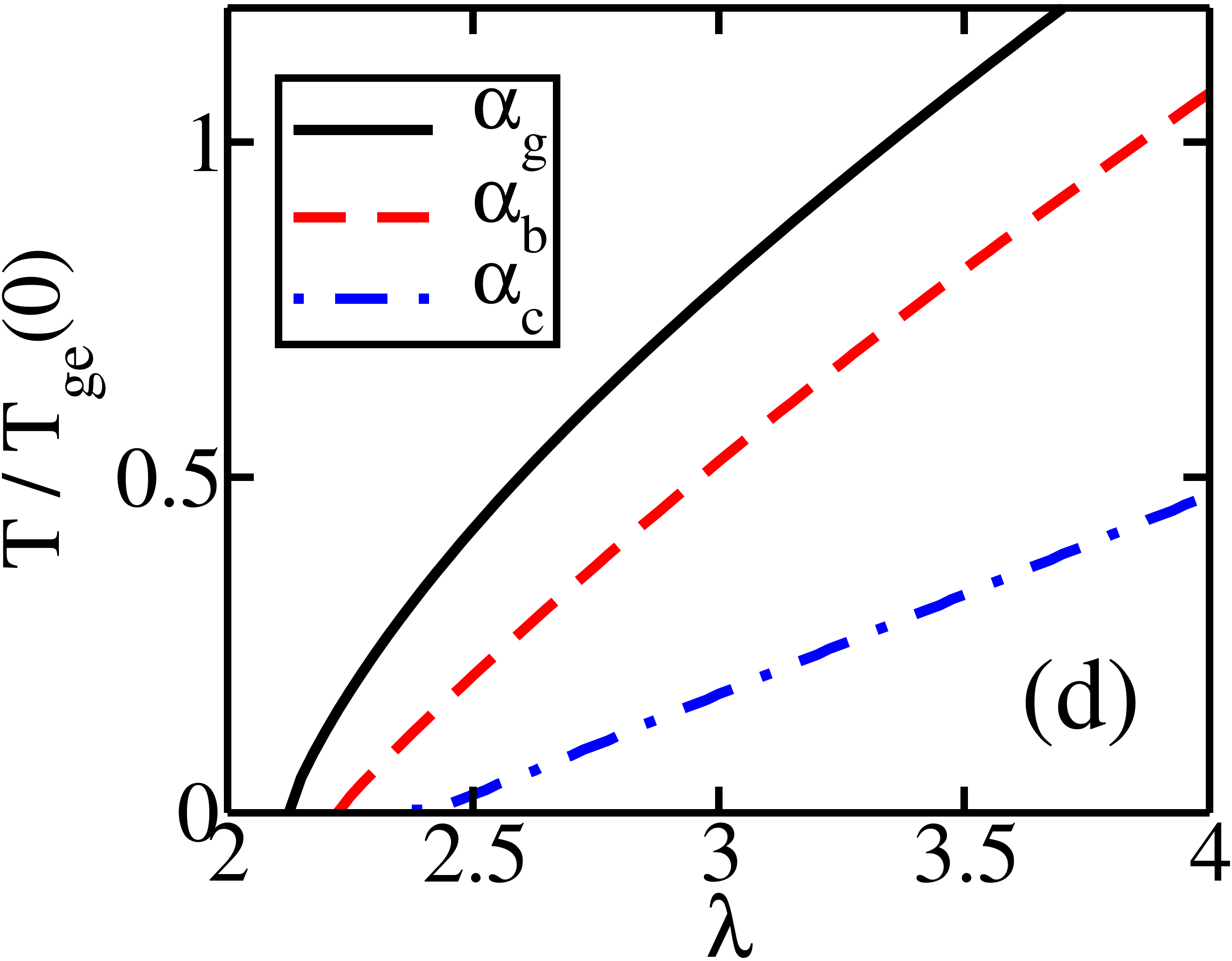}
\caption{(a) Variation of the single particle EE $S_A$ with $\lambda$ at zero temperature for decreasing values of $\alpha$. (b) Surface plot of $S_A$ of bosons with $\alpha=\alpha_g$ showing its variation with temperature $kT$(in units of $J$) and $\lambda$. Solid line: A contour with $S_A=1.1$ which roughly corresponds to the crossover temperature and distinguish different phases of Bose gas. Here three regions I, II and III represents the superfluid phase, localized phase and disordered thermal Bose gas respectively. (c) The same isoentropic contour in the $T/T_{ge}(0)-\lambda$ plane for different values of $\alpha$. (d) The contour with IPR$=0.5$ in the $T/T_{ge}(0)-\lambda$ plane for decreasing $\alpha$. For figures (b-d) filling $\nu=0.7$. For all the plots $N_s=144$ for $\alpha=\alpha_g$.}
\label{EE_boson}
\end{figure}

The analysis of above thermodynamic quantities of ideal Bose gas reveals new interesting features associated with two types of crossover phenomena and finite temperature phases of the Bose gas induced by the quasi-periodic potential. Below the critical point $\lambda <2$, both quasi-condensate and superfluidity coexist in 
the superfluid phase(I) which crosses over to thermal gas phase(III) by increasing the temperature. 
On the other hand, in the localized phase (II) (for $\lambda >2$) the existence of quasi-condensate without superfluidity indicates formation of compressible Bose-glass phase in the presence of interaction. With increasing temperature, both the condensate fraction and IPR decreases which leads to another crossover to disordered thermal gas phase(III). The crossover temperatures vanish at the critical point signifying strong quantum fluctuations at the critical coupling $\lambda =2$. An increase in entropy associated with both the crossover phenomena as well as strong fluctuations near the critical point are also apparent from the isoentropic contours of EE. Similar behavior has also been observed in the phase diagram of hard core bosons\cite{nessi}.
The quasi-condensate phase can survive in weakly interacting regime\cite{shlyap_rev}, however a depletion of condensate fraction and superfluidity due to repulsive interaction may reduce the crossover temperature. 
In weakly interacting condensate, SFF vanishes at the coupling strength larger than the critical value $\lambda=2$ since repulsive interaction favors delocalization\cite{sinha}. Also in the localized regime the IPR increases with much slower rate with increasing values of $\lambda$ due to repulsive interaction\cite{sinha}.
Apart from Bose-glass phase, many-body localized phases and disordered Mott like insulating phases can appear in a system of strongly interacting bosons. 
We also observe both the crossover temperature and degree of localization decreases with decreasing values of quasi-periodicity $\alpha$.

\section{Persistent current of fermions in quasi-periodic potential}\label{sec4} 
In this section, we discuss the transport properties of non-interacting spinless fermions at finite temperatures in the presence of the quasi-periodic potential. Due to the self-duality of the AA model the localization of Fermi energy occurs at the critical coupling $\lambda = 2$. 
Similar to the superflow of the condensate, a current in the fermionic system can also be generated by applying a phase twist at the boundary. With periodic boundary condition this is equivalent to attaching a flux to the fermions moving in a ring. In mesoscopic quantum ring  a persistent current of electrons can be produced by applying a magnetic flux $\phi$ inside the ring. 
In quantum ring the current-flux relationship can depend on many factors such as band structure, disorder, interaction between particles, ring geometry and temperature. In this work, we mainly focus on the current-flux relationship and its variation with the strength of AA potential $\lambda$ and temperature.
\subsection{\bf Persistent current at zero temperature}\label{sec4.1}
We start by reviewing the persistent current at zero temperature. In order to calculate the persistent current one considers a phase-twisted Hamiltonian for fermions similar to what's shown in Eq.\ \ref{ham_twist} with $\theta=2 \pi \frac{\phi}{\phi_0}$, where $\phi_0=h/e$ is the unit flux quanta and operators ${a_l},{a_l}^\dagger$ follow fermionic anti-commutation relation. After diagonalization of this Hamiltonian, the single particle energy levels $\epsilon_{n}(\phi)$ are obtained to calculate the persistent current, which is given by  \cite{Imry,cheung1}, 
\begin{equation}
I_c(\phi)=-\frac{\partial E_0}{\partial \phi} 
\label{Ic}
\end{equation}
where $E_0 = \sum_{n}\epsilon_{n}(\phi) \theta(E_{F} - \epsilon_{n})$ is the ground state energy of the system and $E_{F}$ is the Fermi energy at zero temperature. 
In absence of any potential, the energy dispersion is given by $\epsilon_n(\phi)=-2J\cos(\frac{2\pi}{N_s}(n+\frac{\phi}{\phi_0}))$ where $-{N_s}/2 \leq n < {N_s}/2$. For $N$ fermions in $N_s$ sites, the persistent current can be written as\cite{cheung1}, 
\begin{equation}
I_c=-I_0 \frac{\sin (\frac{\pi}{N_s}(2\frac{\phi}{{\phi}_0}  + \eta))}{\sin(\frac{\pi}{N_s})} 
\label{pc}
\end{equation}
where $I_0=\frac{4 \pi J}{N_s {\phi}_0}\sin(N \pi/N_s)$. The persistent current $I_c$ exhibits periodic variation with flux $\phi/\phi_0$ and a phase shift $\eta$ is generated due to the parity of the number of fermions $N$.
For odd $N$, $\eta=0$ in region $-0.5\leq\frac{\phi}{{\phi}_0}<0.5$ and  for even $N$, $\eta=-1$ in region  $0\leq\frac{\phi}{{\phi}_0}<1$.
Hereafter, we scale the current $I_c$ by $I_0$ which depends on the filling of the fermions and we mainly consider half filled system. 

We now discuss the effect of the quasi-periodic potential on the persistent current at zero temperature to study the localization transition. We calculate the persistent current in AA potential using Eq.\ \ref{Ic} to investigate its variations with the strength of the potential and with the applied flux.
\begin{figure}
\centering
\includegraphics[width=5.15cm,height=4.6cm]{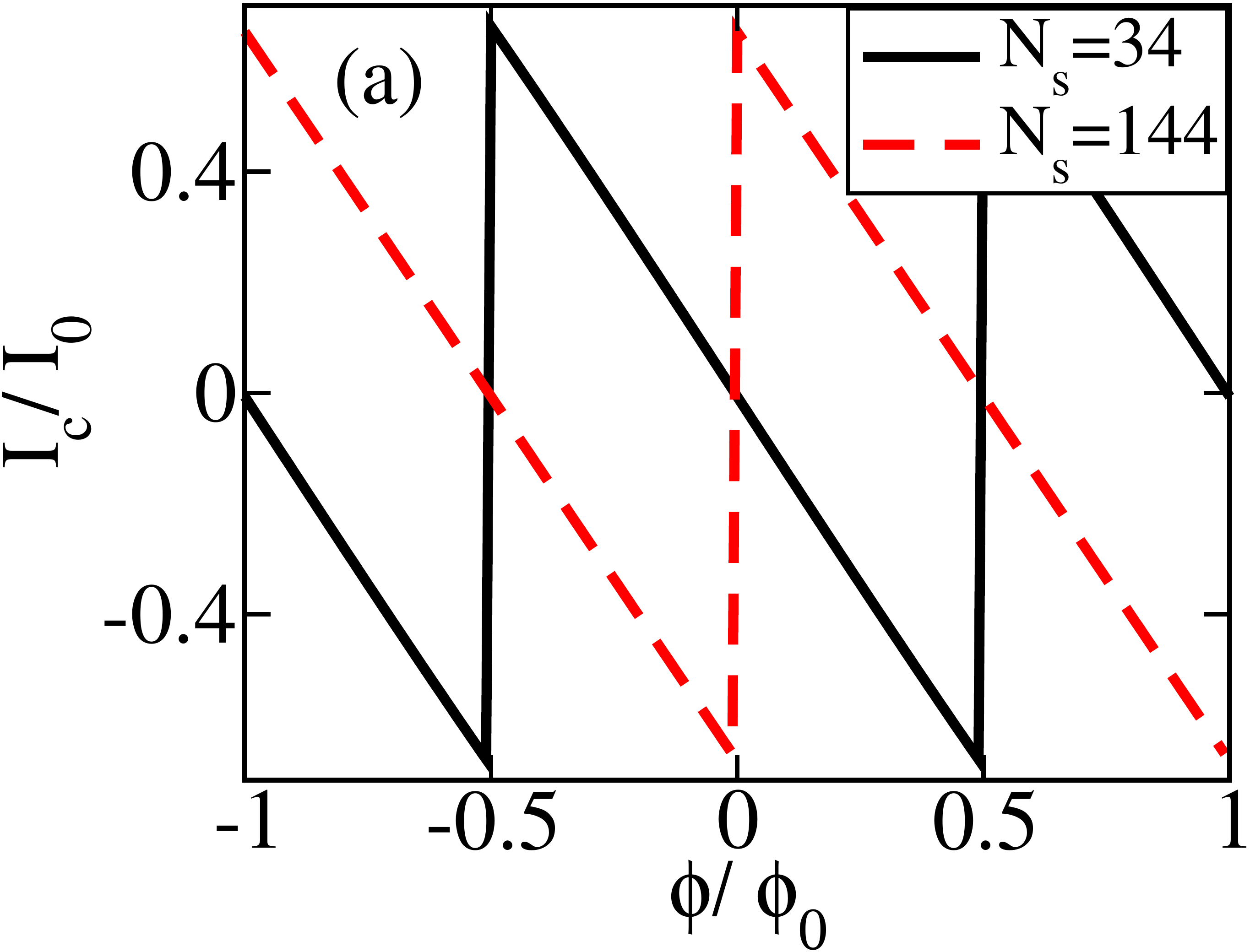}
\includegraphics[width=5.15cm,height=4.6cm]{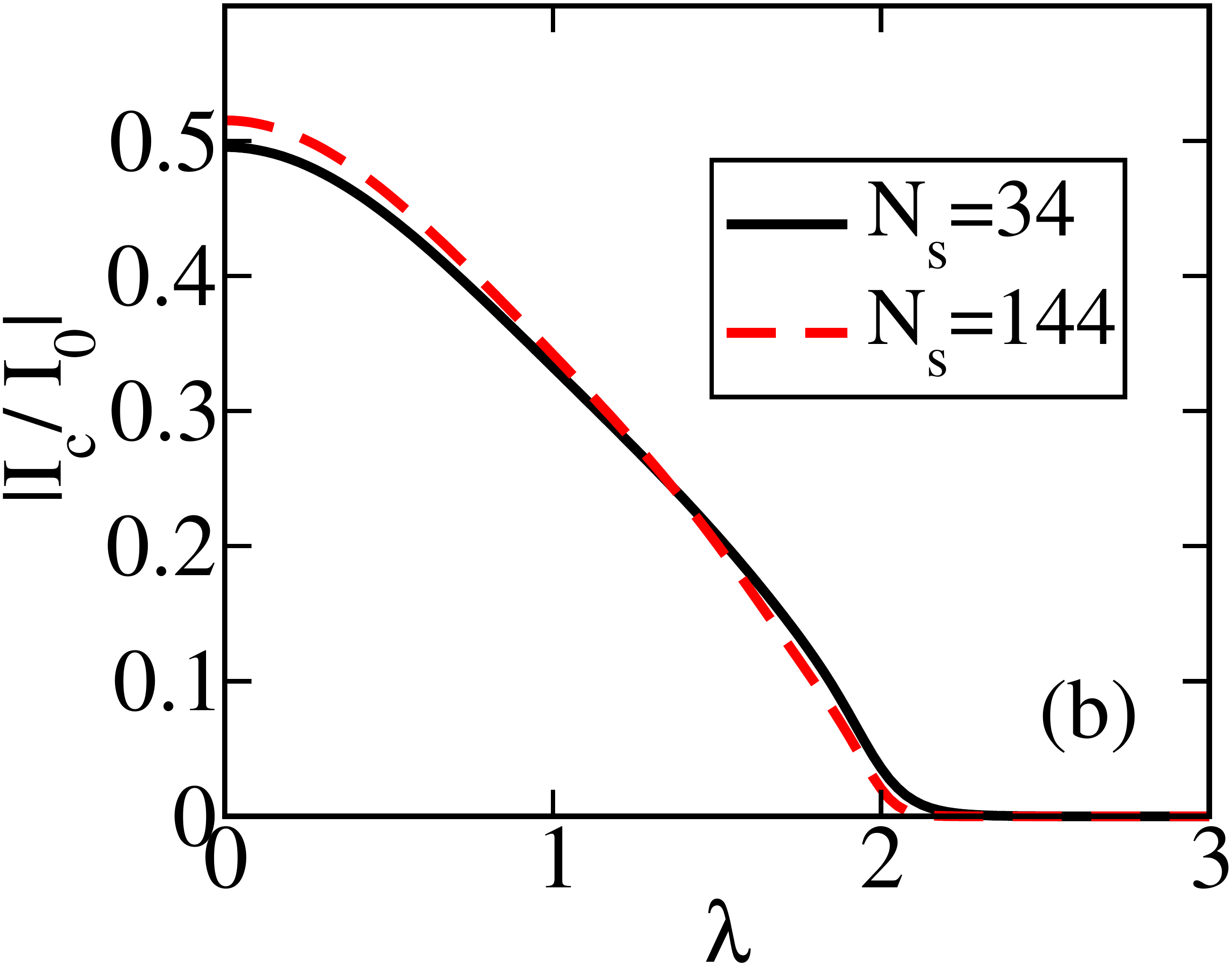}
\includegraphics[width=5.3cm,height=5.05cm]{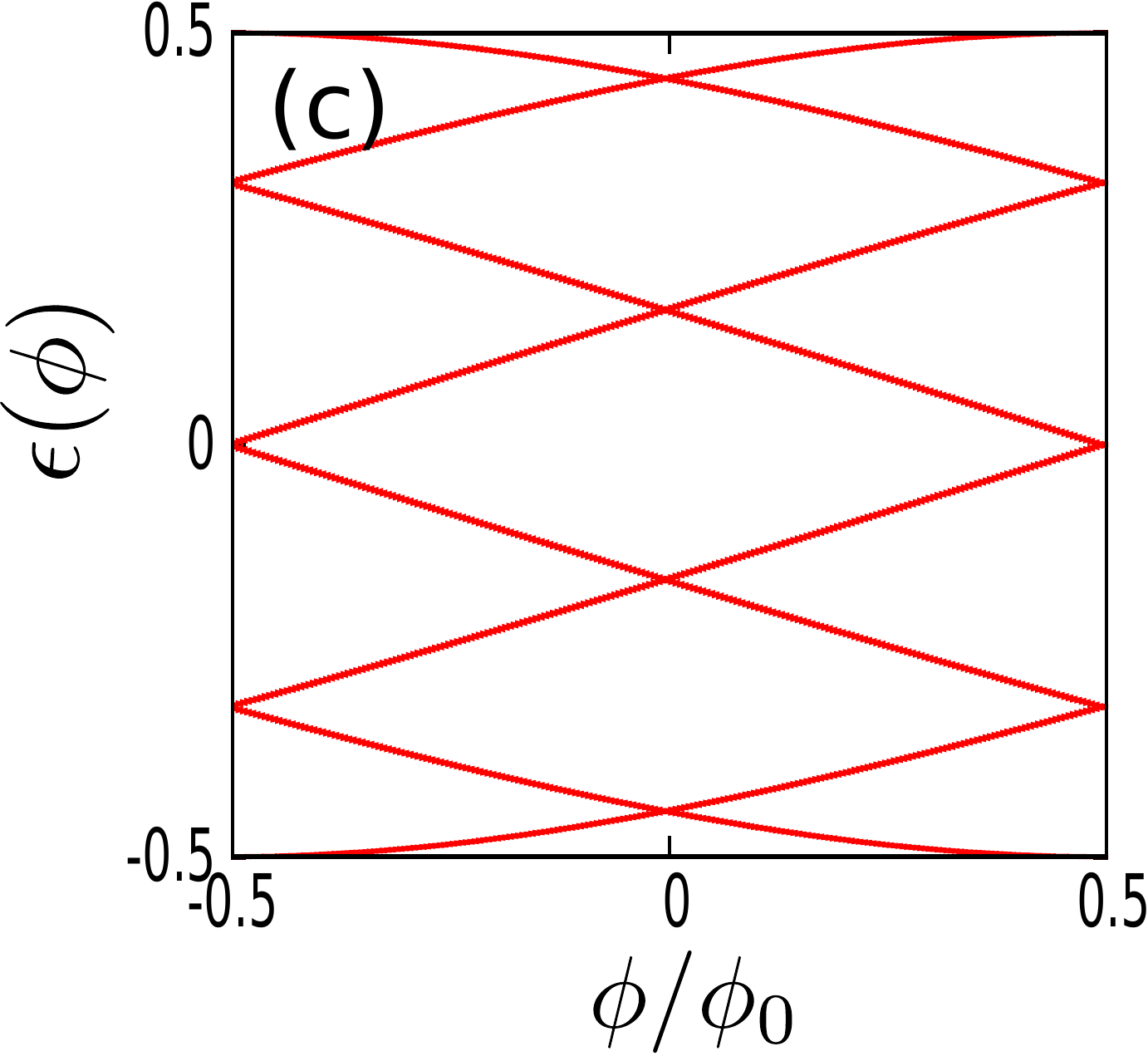}
\includegraphics[width=5.05cm,height=5.05cm]{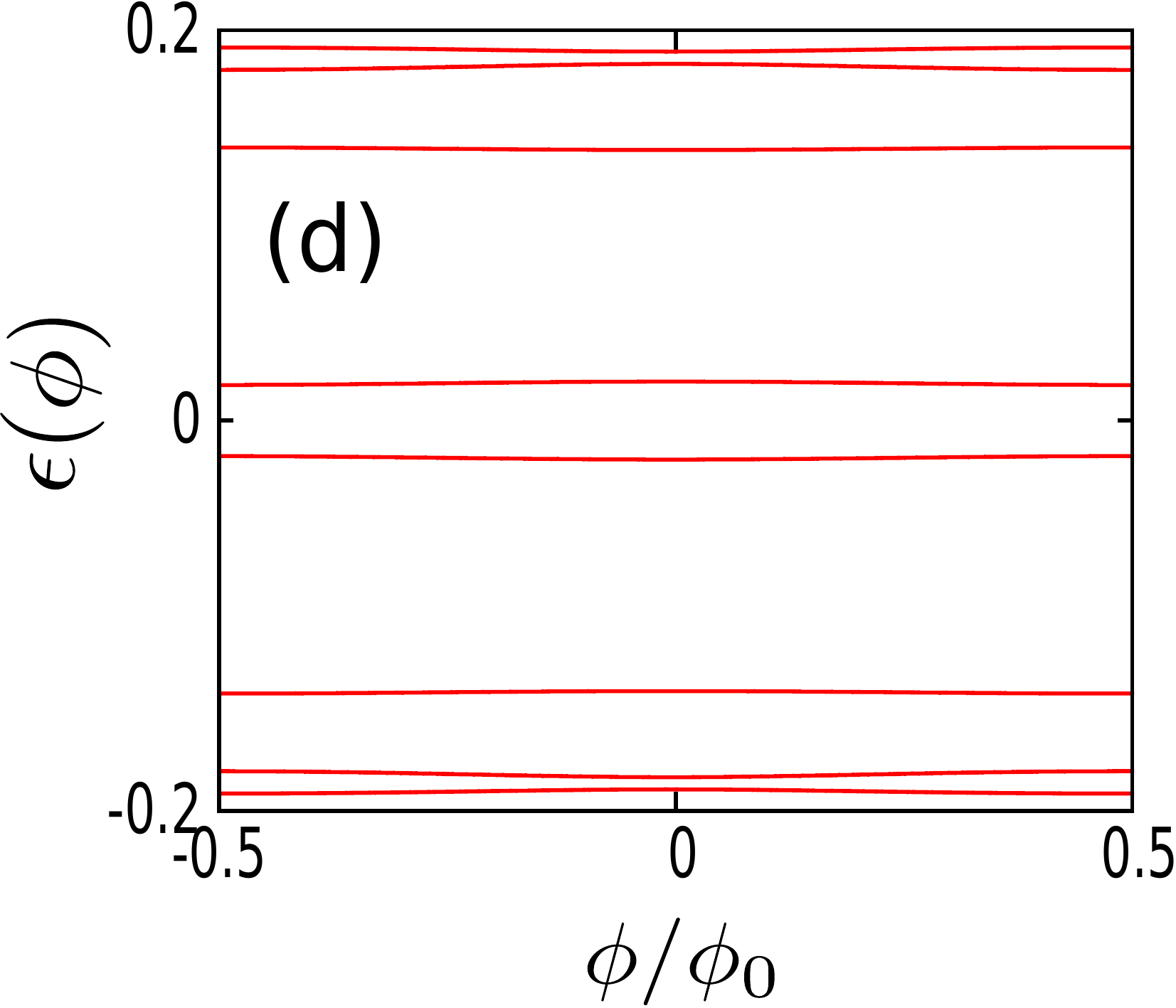} 

\caption{(a) Variation of the dimensionless current with the applied flux at zero temperature for $\lambda=1.0$ and two different values of $N_s=34$ and $N_s=144$ with odd and even number of fermions respectively at half-filling. (b) Decay of the amplitude of the current with $\lambda$ at constant $\phi=-0.25\phi_0$ for half-filled fermions in two different $N_s$. Variation of a few midband energy levels with flux for (c) $\lambda=0.5$ and (d) $\lambda=2.1$ respectively for $N_s=34$. For all plots $\alpha=\alpha_g$.}
\label{pc_zeroT}
\end{figure} 
The oscillatory behavior of the current flux relationship for two different system size with even and odd number of fermions at half filling are shown in Fig.\ \ref{pc_zeroT}(a). 
In both cases the persistent current shows very sharp sawtooth like periodic oscillations with flux $\phi/\phi_0$. It is important to note that although the shape and periodicity of the oscillations of $I_c/I_0$ are same for two different system size, the oscillations are shifted by an amount $0.5$ in $\phi/\phi_0$ due to different values of $\eta$ for even and odd number of fermions. 
The localization effect due to the AA potential is evident from the decay of the amplitude of the current. The amplitude of the persistent current $|I_c/I_0|$ decreases with increasing coupling strength $\lambda$ and vanishes in the localized regime where $\lambda>2$, which is shown in Fig.\ \ref{pc_zeroT}(b) at constant $\phi=-0.25\phi_0$ for two different number of sites $N_s=34$ and $N_s=144$.
Here we point out that although the amplitude of the scaled current $I_c/I_0$ for two different system size are almost same (see Fig.\ \ref{pc_zeroT}(b), the magnitude of $I_c$ is smaller for larger sites as $I_0\sim1/N_s$. 
To qualitatively understand the vanishing of the current in localized phase for half-filling, we investigate the behavior of a few midband energy levels $\epsilon(\phi)$ which are shown in Fig.\ \ref{pc_zeroT}(c-d). In the delocalized phase the energy levels show a strong variation with $\phi$ whereas in the localized regime they form almost flat band which are independent of $\phi$. This leads to the vanishing of current $I_c$ in the localized regime in accordance with Eq.\ \ref{Ic}.
\subsection{\bf Persistent current at finite temperature}\label{sec4.2}
Next we discuss the effect of temperature on the current. In presence of finite temperature the persistent current can be calculated in grand canonical and canonical ensembles using the definitions, which are connected by the following thermodynamic relation \cite{imry}
\begin{equation}
{\bigg(\frac{\partial \Omega}{\partial \phi}\bigg)}_\mu ={\bigg (\frac{\partial F}{\partial \phi}\bigg )}_N = -{I_c}^\beta,
\end{equation}
where $\Omega$ is the grand potential and $F$ is the Helmholtz free energy \cite{pathria}. We first consider the zero disorder case. As long as the temperature is kept well below the level spacing $\Delta_F$ at the Fermi energy, the current remains unaffected by temperature and it starts deviating from the zero temperature behavior 
when $kT\sim\Delta_F$ \cite{cheung1} (see Fig.\ \ref{pc_T}(a) and (b)).
As seen from Fig.\ \ref{pc_T}(a) and (b), the temperature has two main effects on the 
current $I_c$; very sharp sawtooth like oscillations of $I_c/I_0$ with flux are smoothened out and its amplitude decreases with increasing temperature keeping the period of the oscillation same. 
At even higher temperatures, the persistent current falls off substantially and vanishes at a certain temperature depending on the value of $\lambda$, which is denoted by $T^{*}(\lambda)$. 
At finite temperatures, the variation of the amplitude of the persistent current $I^{\beta}_c/I_0$ with increasing strength of AA potential $\lambda$ is shown in Fig.\ \ref{Tstar}(a), for a fixed value of flux $\phi=-0.25\phi_0$. When $kT \leq \Delta_F$, the persistent current decreases with increasing $\lambda$ and vanishes at the critical value $\lambda = 2$ confirming the localization at zero temperature. For higher temperatures the amplitude of the current decreases and eventually vanishes at smaller values of $\lambda$. It is evident from Fig.\ \ref{pc_T}(b) and Fig.\ \ref{Tstar}(a) that both thermal effect and quasi-periodic potential destroy the persistent current in a ring. 
From the vanishing of the current $I^{\beta}_c$ we calculate the temperature $T^{*}(\lambda)$ which characterize the localization transition of the fermions at finite temperature. To eliminate the size dependence, variation of the dimensionless scaled temperature $t^{*}(\lambda) = T^{*}(\lambda)/T^{*}(0)$ with $\lambda$ is presented in Fig.\ \ref{Tstar}(b) for different values of quasi-periodicity $\alpha$ of the AA-potential. This dimensionless temperature $t^{*}(\lambda)$ decreases with increasing coupling strength $\lambda$ and vanishes in the localized regime $\lambda >2$. It is interesting to note that unlike the crossover temperatures of bosons, $t^{*}(\lambda)$ exhibits non-monotonic behavior with $\alpha$.
\begin{figure}
\centering
\includegraphics[width=5.26cm,height=4.7cm]{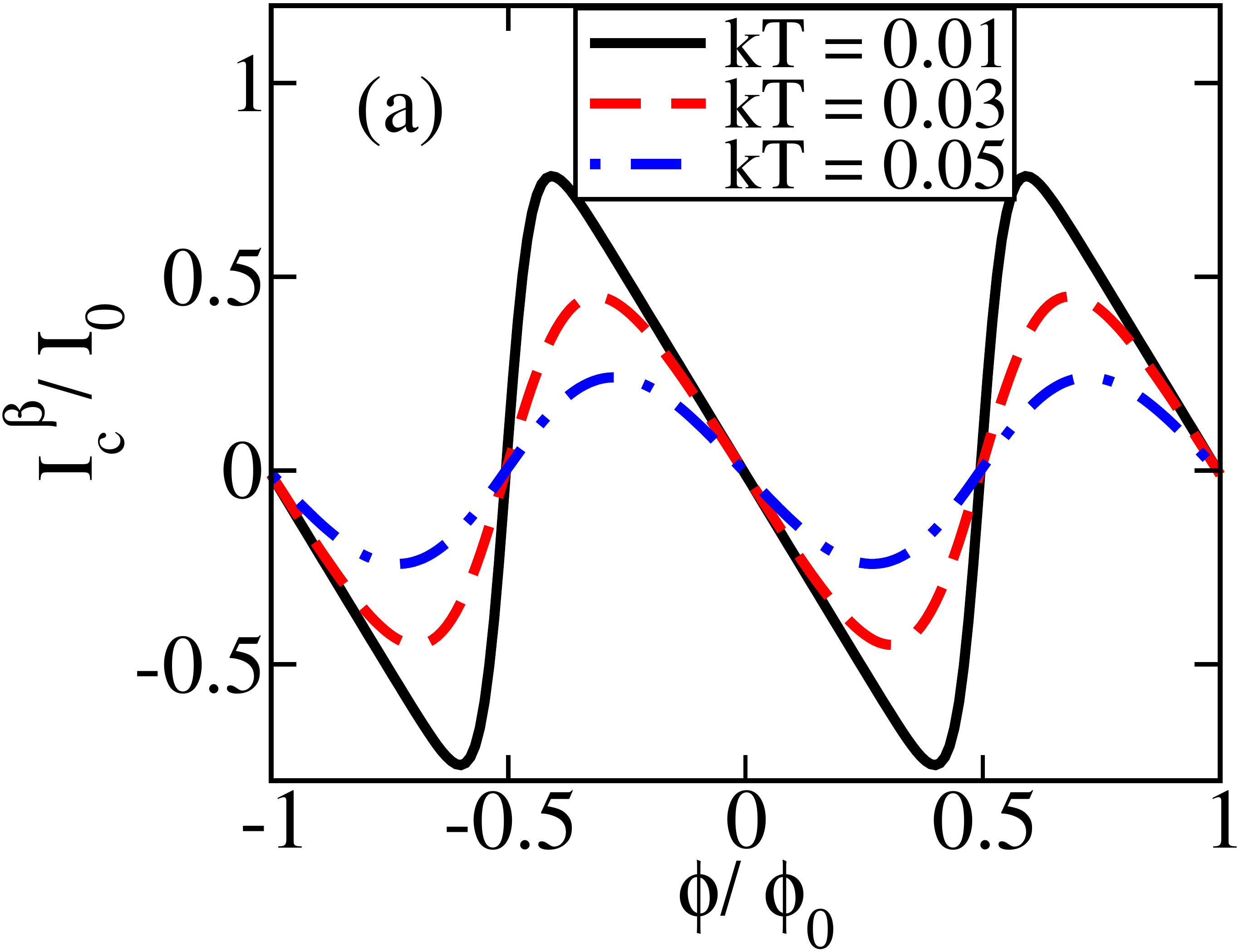}
\includegraphics[width=5.26cm,height=4.7cm]{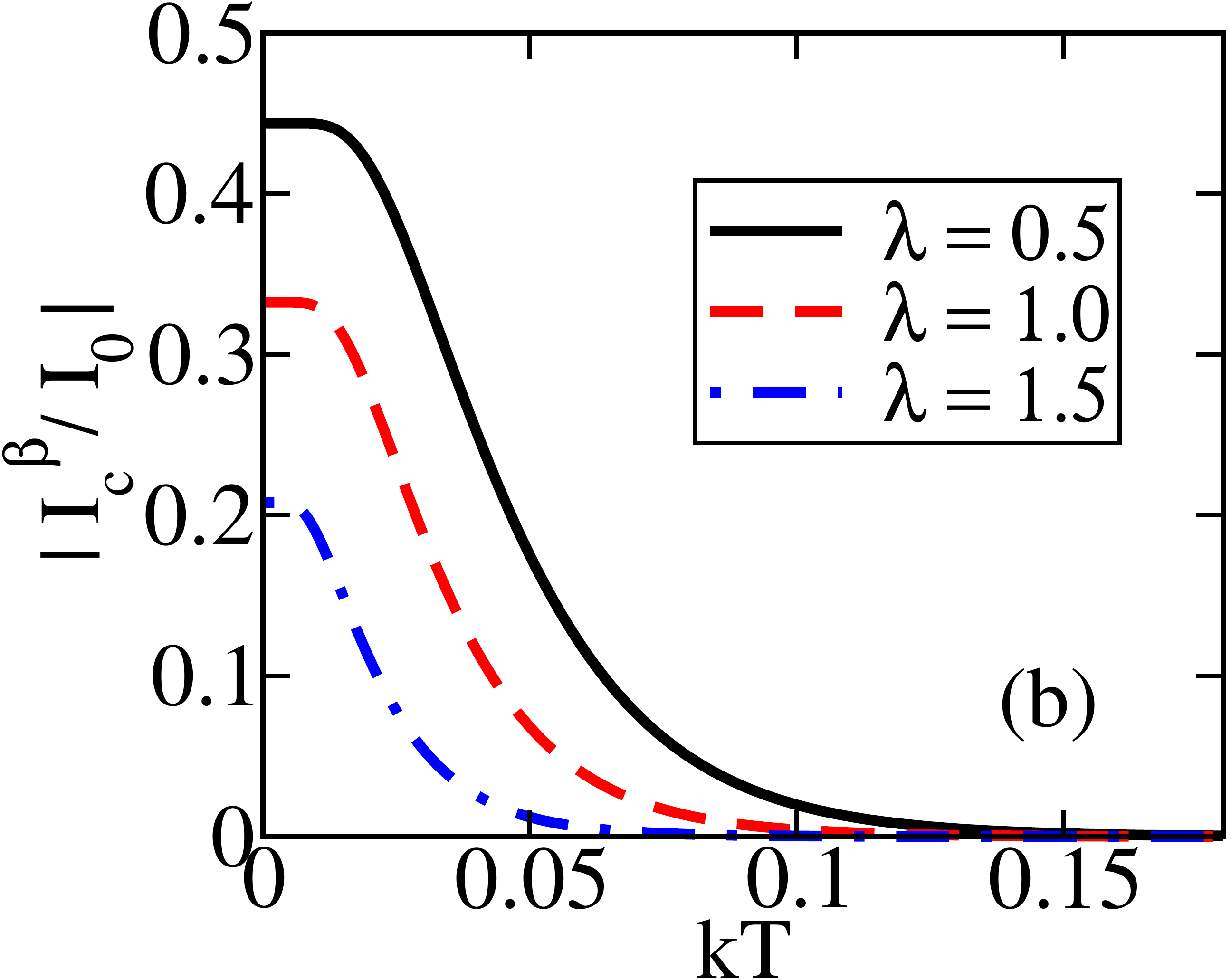}
\caption{(a) The dimensionless current-flux relationship for increasing temperature $kT$(in units of $J$) at fixed strength of AA potential $\lambda=1.0$. (b) Decay of the amplitude of the dimensionless current $|I^{\beta}_{c}/I_0|$ with temperature $kT$ (in units of $J$) for different values of $\lambda$ at constant $\phi=-0.25\phi_0$. For both the plots $N_s=34$, $N/N_s = 0.5$ and $\alpha=\alpha_g$.}
\label{pc_T}
\end{figure} 
This indicates that the energy gap at the Fermi energy has non trivial functional dependence on quasi-periodicity $\alpha$ of AA model, since the temperature dependence of the persistent current is mainly controlled by the energy gap $\Delta_F$.
This non monotonic behavior of persistent current with quasi-periodicity of lattice is a new effect that can have relevance in mesoscopic systems and testable in experiment.  
%
%
The localization transition of fermions as indicated by vanishing of persistent current always occurs at the self dual critical point $\lambda = 2$ irrespective of Fermi energy due to the absence of mobility edge in AA model. 
In general, duality of AA model can not be restored in the presence of interaction and mobility edge can appear, as a result the localization transition can show interesting dependence on Fermi energy.
\begin{figure}
\centering
\includegraphics[width=5.26cm,height=4.7cm]{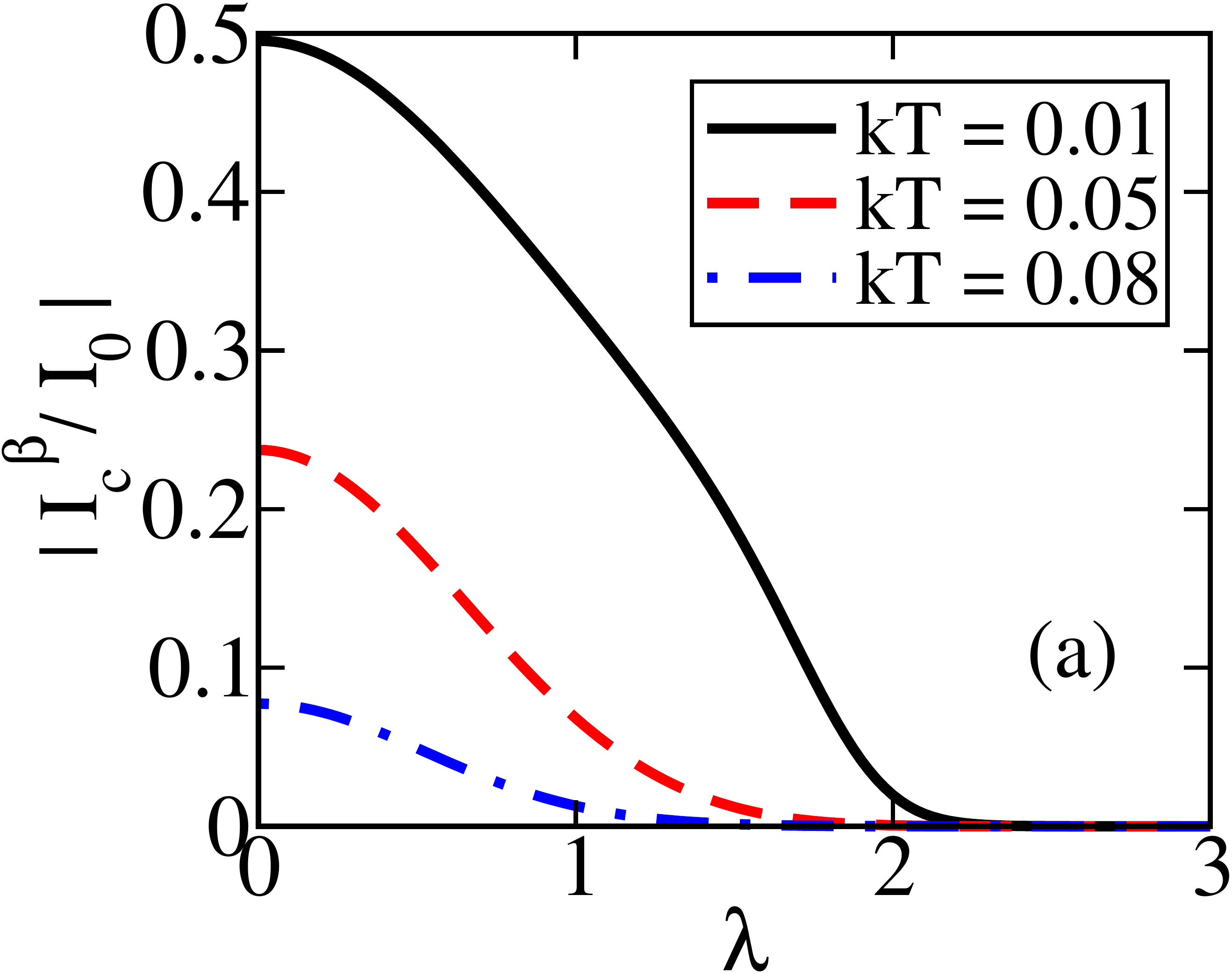}
\includegraphics[width=5.26cm,height=4.7cm]{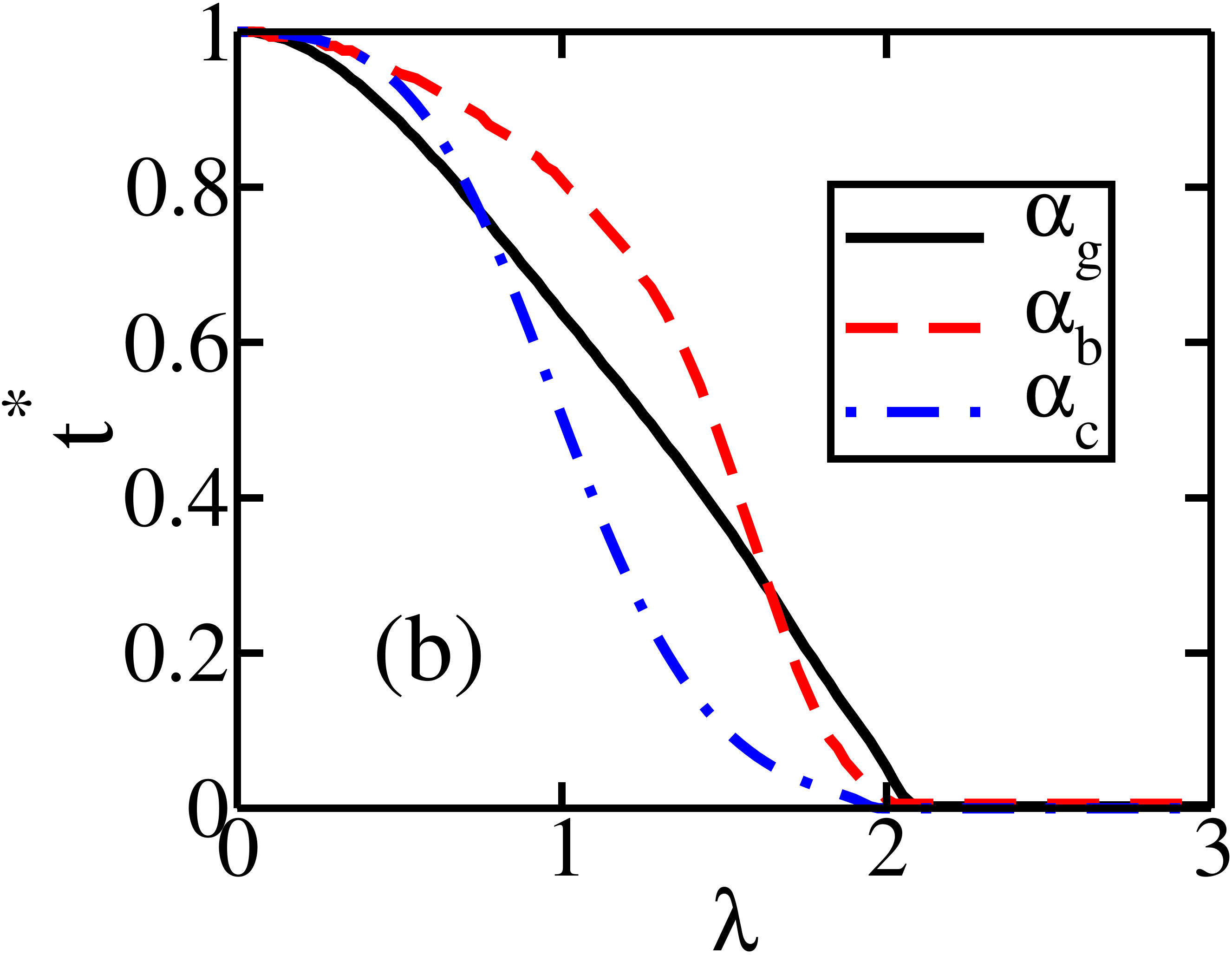}
\caption{(a) Dependence of the current $I^{\beta}_c$ (in units of $I_0$) with $\lambda$ for different temperatures $kT$ (in units of $J$) for $N_s=34$ and $\alpha=\alpha_g$.
(b) The dimensionless scaled temperature $t^*$ as a function of $\lambda$ for different values of $\alpha$. Here $N_s=144$ for $\alpha=\alpha_g$. For both the plots $N/N_s = 0.5$, $\phi=-0.25\phi_0$}.
\label{Tstar}
\end{figure}

\section{Conclusion}\label{sec5}

To summarize, in this work we study the thermodynamics of ideal Bose gas as well as the transport properties of non interacting bosons and fermions in the
AA potential to investigate the localization phenomena at finite temperatures.

The formation of quasi-condensate phase, superfluidity and localization phenomena of ideal Bose gas at finite temperature are studied which reveals new effects and two crossover phenomena due to the presence of quasi-periodic disorder.
The crossover temperature of quasi-condensate phase is calculated from the ground state number fluctuations in both canonical and grand canonical ensembles. 
For $\lambda <2$, the quasi-condensate phase and superfluidity coexist and both the phases change to thermal gas phase following a smooth crossover by increasing the temperature. 
The crossover temperature decreases with increasing strength of the AA potential $\lambda$ and vanishes at the critical strength $\lambda_c =2$ signifying the enhanced fluctuation at the critical coupling. 
Although the crossover temperature is a system size dependent non universal quantity, 
appropriate scaling of it reveals `ensemble equivalence' as well as its universal variation with $\lambda$, except in close vicinity of the critical point where the finite size effect is relevant. 
Interestingly, the scaled energy gap of the ground state also shows the same universal variation with $\lambda$.
The scaled crossover temperature decreases with decreasing quasi-periodicity $\alpha$ of the AA model and vanishes at the critical point following the power law $t_{ce} \sim |\lambda - \lambda_c|^{\gamma(\alpha)}$, where the exponents $\gamma(\alpha)$ depends on quasi-periodicity. Similar to the quantum critical point, this behavior reflects enhanced quantum fluctuation due to the self dual critical point for localization transition in AA model. 
In the localized regime for $\lambda > 2$, the quasi-condensate phase can still survive in absence of superfluidity and the crossover temperature increases with $\lambda$ indicating possible formation of Bose-glass phase at finite temperature in the presence of interaction. With increasing temperature both condensate fraction and IPR decreases and the localized quasi-condensate crosses over to disordered thermal gas phase. These crossover phenomena and fluctuations near the critical coupling can also be identified from enhancement of entanglement entropy. 
It is interesting to note that the scenarios observed in ideal Bose gas can qualitatively capture the essential features the phase diagram of hardcore bosons.

To investigate the localization of the Fermi gas at finite temperature we have studied
the persistent current of the fermions in the presence of the AA potential. By increasing the strength of the AA potential the decay of the persistent current indicates localization at the Fermi energy. 
The persistent current also exhibits periodic oscillations with the variation of the flux (phase twist) and the amplitude of the current decreases with increasing the disorder strength $\lambda$ as well with the increasing temperature.
Unlike the bosonic case, the temperature corresponding to the vanishing of persistent current of fermions shows a non monotonic dependence on the quasi-periodicity $\alpha$ of the AA potential.
The effect of interaction in non-ideal realistic systems can give rise to new effects and new phases which requires further analysis. In the presence of interaction
`duality' of the AA model will be violated and mobility edge can appear which can have interesting consequences on localization transition at finite temperatures.

The universal behavior of the crossover temperature to quasi-condensate phase, finite temperature transport properties and enhanced fluctuations near the self dual critical point
of the AA model are the main results of the present work which can have relevance in the future experiments on ultracold quantum gases in the presence of a bichromatic optical lattice.

\section*{Acknowledgements} 
We thank Sayak Ray for helpful discussions. NR thanks UGC-CSIR, India for doctoral fellowship and acknowledges support from IISER-K, where the work has been started.

\end{document}